\documentclass[pra,twoside,fleqn,floatfix,showpacs,preprint]{revtex4}
 \usepackage{graphicx,times,amsmath,mathrsfs,amsfonts,amssymb,units}
 \usepackage[usenames]{color}

\newcommand{\Eloss}{\Delta E_{\mathrm{loss}}}
\newcommand{\Hop}{\ensuremath{\mathscr{H}}}
\newcommand{\vFind}{{\bf F}_{\mathrm{ind}}}
\newcommand{\FTPhiind}{\widehat{\Phi}_{\mathrm{ind}}}
\def\charge{{{e}}}
\def\exponent{{\mathrm{e}}}
\def\deltan{\delta n}
\def\deltaFn{\delta\widetilde{n}}
\def\qb{{\bf q}}
\def\Kb{{\bf K}}
\def\Vpb{{\bf V}_\parallel}
\def\rb{{\bf r}}
\def\rbs{{\bf r}_s}
\def\xb{{\bf x}}

\def\grad{\nabla_{\!\!s}}
\def\vb{{\bf u}}
\def\dxdt{{\bf \dot{\bf x}}}

\def\bigoh{\mathcal{O}}
\def\dotxi{{\bf \dot{\xi}}}
\def\ddotxi{{\bf \ddot{\xi}}}
\def\Area{\mathcal{A}}
\def\xfb{\widetilde{\xi}}

\def\Phiind{\Phi_{\text{ind}}}

\def\Vext{V_{\text{ext}}}

\def\vareps{\varepsilon}
\def\VV{\breve{V}_{\text{ext}}}
\begin{document}

\title{Plasmon excitations on a single-wall carbon nanotube by external charges: two-dimensional, two-fluid hydrodynamic model}

\author{D. J. Mowbray$^{\text{1,2}}$}

\author{S. Segui$^{\text{3}}$}
\altaffiliation{Also at Consejo Nacional de Investigaciones Cient\'{\i}ficas y T\'ecnicas (CONICET), Argentina.}

\author{J. Gervasoni$^{\text{3}*}$}

\author{Z. L. Mi{\v{s}}kovi{\'{c}}$^{\text{2}}$}
\email[Corresponding author: ]{zmiskovi@uwaterloo.ca}

\author{N. R. Arista$^{\text{3}}$}

\affiliation{ $^{\text{1}}$Nano-Bio Spectroscopy group and ETSF Scientific Development Centre, Dpto.~F\'{\i}sica de Materiales,
Universidad del Pa\'{i}s Vasco,
Centro de F\'{\i}sica de Materiales CSIC-UPV/EHU- MPC and DIPC, Av.~Tolosa 72, E-20018 San Sebasti{\'{a}}n, Spain\\
$^{\text{2}}$Department of Applied Mathematics, University of Waterloo,\\ Waterloo, Ontario, Canada N2L 3G1\\
$^{\text{3}}$Centro At\'{o}mico Bariloche, Comisi\'{o}n Nacional de Energ\'{i}a At\'{o}mica, Avenida Bustillo 9500, 8400 S.C. de
Bariloche, R\'\i o Negro, Argentina }

\date{\today}
\begin{abstract}
We present a quantization of the hydrodynamic model to describe the excitation of plasmons in a single-walled carbon nanotube by
a fast point charge moving near its surface at an arbitrary angle of incidence. Using a two-dimensional electron gas represented
by two interacting fluids, which takes into account the different nature of the \(\sigma\) and \(\pi\) electrons, we obtain
plasmon energies in near-quantitative agreement with experiment. Further, the implemented quantization procedure allows us to
study the probability of exciting various plasmon modes, as well as the stopping force and energy loss spectra of the incident
particle.\end{abstract}


\maketitle

\section{Introduction}

Plasmon excitations in carbon nanotubes (CNTs) continue to attract attention for a variety of applications, e.g., in the context
of their optical response \cite{Shuba_2009,Nakanishi_2009,Lidorikis_2009}. On the other hand, plasmon excitations are most
effectively probed by fast-moving charged particles, such as in electron energy loss spectroscopy (EELS) using a scanning
transmission electron microscope (STEM). This technique has proven to be a powerful tool for investigating the dynamic response
of CNTs \cite{Pichler_1998,Stephan_2002,Kramberger_2008,Kramberger_2008b} and, more recently, graphene
\cite{Eberlein_2008,Liu_2008}. Moreover, plasmon excitations were observed to play an important role in electronic excitations
of CNTs exposed to ion bombardment \cite{Krasheninnikov}. On the theoretical side, in addition to \emph{ab initio} calculations
\cite{Marianopoulos_2003,Kramberger_2008}, simpler models have been also used to study plasmon excitations in carbon
nano-structures, such as the empirical dielectric tensor \cite{Taverna_2002} and the hydrodynamic model \cite{Wang_1996}.

Since carbon nano-structures present physical realizations of one-atom thick layers of an electron gas, it is no surprise that a
two-dimensional (2D) version of the hydrodynamic model was used early on in studying plasmon excitation in CNTs
\cite{Longe_1993,Yannouleas_1994,Yannouleas_1996}. We note that a planar 2D hydrodynamic model of the electron gas was pioneered
by Fetter \cite{Fetter_1973,Fetter_1974} in 1973. This model has subsequently been used to gain qualitative understanding of
plasmon excitations in various quasi-2D electronic structures, including semiconductor inversion layers, quantum wells, and thin
metallic films. Such a model assumes that all electrons belong to a single fluid characterized by two parameters only, the
equilibrium surface density $n_0$ and the effective electron mass $m_*$. Owing to its simplicity and versatility in handling
difficult geometric constraints, multilayered structures, and the presence of dielectric environment, the single-fluid version
of the 2D hydrodynamic model has been used in a significant number of applications of CNTs and Fullerene molecules
\cite{Stockli_2001,Wang_2004,Wei_2004,Mowbray_2005,Tuktarov_2005,Zhou_2006,Borka_2006,Miano_2006,Javaherian_2009}.

However, treating the four valence electrons per atom in carbon nano-structures as belonging to the same fluid fails to capture
significant differences in their bonding. Three of these valence electrons are involved in the strong $\sigma$ bonds
characterizing the $sp^2$ hybridization within a layer of carbon atoms. At the same time, one electron occupies a weakly bound
$\pi$ orbital that is largely responsible for the conductivity properties of carbon nano-structures. While the $\pi$ electron
bands make graphene a zero-gap semiconductor, the so-called graphene nano-ribbons (GNRs), as well as CNTs may exhibit metallic
or semiconducting character depending on the symmetry of their underlying atomic structure. On the other hand, the $\sigma$
electron bands exhibit a large gap that gives rise to a characteristic high-frequency feature in the absorption spectra in
various allotropic forms of carbon.

Consequently, inspired by the empirical model of Cazaux \cite{Cazaux_1970}, Barton and Eberlein \cite{Barton_1993} proposed a
two-fluid version of the 2D hydrodynamic model. Their model treats the $\sigma$ and $\pi$ electrons as two classical fluids with
equilibrium densities $n^0_\sigma=3\,n_{at}$ and $n^0_\pi=\,n_{at}$ (where $n_{at}\approx 38$ nm$^{-2}$ is graphene's atomic
density), respectively, which they superimposed on the surface of a C$_{60}$ molecule. By introducing an empirical restoring
frequency for the $\sigma$ electron fluid, the two-fluid 2D hydrodynamic model of Barton and Eberlein was shown to give rise to
two groups of plasmons. This explained well the two dominant absorption features in the frequency ranges of 5-10 eV and 15-30
eV, in both Fullerene molecules \cite{Barton_1993,Gorokhov_1996} and single-wall CNTs (SWCNTs) \cite{Jiang_1996}.

On the other hand, Mowbray \textit{et al.} introduced the quantum, or Fermi pressure in the $\sigma$ and $\pi$ electron fluids,
and reformulated the two-fluid 2D hydrodynamic model without invoking any empirical parameters \cite{Mowbray_2004}. They found
that the electrostatic interaction between the fluids gives rise to two groups of high- and low-frequency plasmons for both
SWCNTs \cite{Mowbray_2004} and multi-wall CNTs (MWCNTs) \cite{Chung_2007} in similar ranges as those obtained by Barton and
Eberlein \cite{Barton_1993}. Versatility of this version of the two-fluid 2D hydrodynamic model was further exploited in
studying the interactions of CNTs with dielectric media \cite{Mowbray_thesis,Mowbray_2006}, as well as in applications to the
propagation of electromagnetic radiation through CNTs \cite{Moradi_2007,Nejati_2009,Moradi_2010}.

In the present paper, we perform a second quantization of the two-fluid 2D hydrodynamic model for electron gas confined on the
surface of an infinitely long cylinder representing a free SWCNT, and use it to analyze the spectra of plasmon excitations
induced by a classical non-relativistic charged particle, passing by the nanotube under an arbitrary angle of incidence. This
approach enables us to study the effects of multiple plasmon excitations by external particle beams, similar to the EELS studies
of thin metallic films on solid surfaces \cite{Sunjic}. Such a study can help elucidate which plasmon modes are  most
effectively excited depending on the incident projectile trajectory. Besides obvious applications in the EELS studies of CNTs,
the results of this work may help to understand the role of electronic channels in energy deposition in carbon nano-structures
during ion irradiation \cite{Krasheninnikov}.

After describing the model and outlining its second quantization in section \ref{II}, we shall discuss our results for the
plasmon excitation spectra, total energy loss, and the stopping force on the external charge in section \ref{III}. Our
concluding remarks will be followed by a description of the statistical properties of plasmon excitations in appendix
\ref{app:a}, a discussion of the relation to the semi-classical model in appendix \ref{app:b}, and an outline of how the results
for planar 2D electron gas may be obtained from the formalism for cylindrical geometry in appendix \ref{app:c}. We shall use the
Gaussian electrostatic units unless otherwise indicated.

\section{Basic Theory}\label{II}

\subsection{Hydrodynamical two-fluid Hamiltonian}

The total Hamiltonian, $\Hop$, for the $\sigma$ and $\pi$ electron fluids can be written as
\begin{widetext}
\begin{eqnarray}
\label{eq:1} \Hop &=& \sum_{\nu=\sigma,\pi} \int\!d^2\rbs \,n_\nu(\rbs,t)\left[
    \frac{1}{2m_\nu^*}\|{\bf p}_\nu(\rbs,t)\|^2+\frac{\kappa_\nu}{2}
\|\xb_\nu(\rbs,t)\|^2 +V(\rbs,t)
\right]\nonumber\\
&+& \sum_{\nu=\sigma,\pi} \int\!d^2\rbs\,n_\nu(\rbs,t)\left\{\frac{\pi
    \hslash^2}{2 m_\nu^*} n_\nu(\rbs,t)
-\charge^2\left[\frac{32}{9\pi}n_\nu(\rbs,t)\right]^{1/2}\right\}
\nonumber\\
&+& \frac{\charge^2}{2}\sum_{\nu=\sigma,\pi}\sum_{\nu'=\sigma,\pi} \iint\!d^2\rbs\,
d^2\rbs'\,\frac{n_\nu(\rbs,t)n_{\nu'}(\rbs',t)}{\|\rbs - \rbs'\|}, \qquad \label{H}
\end{eqnarray}
\end{widetext}
where \(\rbs \equiv \{r=R;\varphi,z\}\) is a position on the surface of a single-wall carbon nanotube of radius $R$ aligned with
the $z$-axis of a cylindrical coordinate system with coordinates $\rb=\{r,\varphi,z\}$, whereas \(n_{\nu}\), ${\bf p}_\nu$,
\(\xb_{\nu}\), $m_\nu^*$, and \(\kappa_{\nu}\) are, respectively, the number density per unit area, the fluid momentum field,
displacement field, effective mass, and the restoring-force constant for the $\nu$th electron fluid, with \(\nu=\pi,\sigma\).
Note that the first term in Eq.~(\ref{H}) may be interpreted as the kinetic energy of a classical fluid $\nu$ moving at the
velocity $\vb_\nu(\rbs,t)\equiv{\bf p}_\nu(\rbs,t)/m_\nu^* =\dxdt_\nu(\rbs,t)$, where the dot stands for time derivative. The
second term represents the restoring (R) effects on the fluid displacement due to electron binding in harmonic approximation,
giving rise to restoring frequency $\omega_{\nu r}\equiv\sqrt{\kappa_\nu/m_\nu^*}$. Further, the potential energy per electron,
$V=V_\text{gr}+\Vext$, consists of the ground state energy due to the positive-ion background, $V_\text{gr}(\rbs)$, and the
energy due to the time-dependent external perturbing potential, \( \Vext(\rbs,t)\). The next two terms represent, respectively,
the Thomas-Fermi (TF) and the Dirac's (D) exchange interactions in the local-density approximation for the electron fluid $\nu$,
whereas the last term in Eq.~(\ref{H}) represents Coulomb interactions among electrons.

A few additional comments on the Hamiltonian in Eq.~(\ref{H}) may be in order. While restoring frequencies $\omega_{\nu r}$ play
the role of empirical parameters that are usually determined by fitting to experimental data
\cite{Cazaux_1970,Barton_1993,Stephan_2002,Taverna_2002}, their physical motivation is to take into account the effects of the
electronic band structure of carbon nano-structures in a manner analogous to that invoked in devising the Drude-Lorentz
dielectric function for carbon materials \cite{Calliari}. So, invoking the high mobility of the $\pi$ electrons in metallic
nanotubes and considering that the $\sigma$ electron bands exhibit a gap of about 12 eV, we adopt the restoring frequencies of
$\omega_{\pi r}=0$ eV and $\omega_{\sigma r}=16$ eV, which have been used by previous authors
\cite{Barton_1993,Gorokhov_1996,Jiang_1996}. Next, we note that the TF and D terms are written in the Hamiltonian in
Eq.~(\ref{H}) in a form commensurate with a \emph{planar} 2D electron gas. This may be justified by considering a quasi-free 2D
electron gas with the number density per unit area $n$, occupying the surface of a cylinder with radius $R$, where curvature
effects were found negligible when $Rk_F\gg 1$, where $k_F=\sqrt{2\pi n}$ is the Fermi wavenumber of the corresponding planar 2D
electron gas \cite{Mowbray_thesis,Sato}. Finally, we note that a Hamiltonian, which is similar in form to the one in
Eq.~(\ref{H}), was used by van Zyl and Zaremba \cite{Zyl_1999} who included the so-called von Weizs\"{a}cker, or
gradient-correction term of the form
\begin{eqnarray}
\Hop_\mathrm{vW} = \sum_{\nu=\sigma,\pi} \lambda_\mathrm{vW}\frac{\hslash^2}{8 m_\nu^*}\int\!d^2\rbs\,\frac{\|\nabla
n_\nu(\rbs,t)\|^2}{n_\nu(\rbs,t)}. \label{vW}
\end{eqnarray}
This provides an approximate correction for the non-local effects in the TF interaction in a planar 2D electron gas. However,
there is considerable uncertainty regarding the exact form of such a term. In fact, it was confirmed recently that the von
Weizs\"{a}cker factor $\lambda_\mathrm{vW}$ in Eq.~(\ref{vW}) should vanish in a strict planar case \cite{Salasnich_2007}.
Nevertheless, it has become quite common to use the von Weizs\"{a}cker correction with $\lambda_\mathrm{vW}=1$ in many recent
studies of CNTs \cite{Wang_2004,Wei_2004,Mowbray_2004,Moradi_2007}, even though the effects of curvature on this term are not
known at present. Fortunately, the von Weizs\"{a}cker correction was found to only affect plasmon dispersion at very short
wavelengths \cite{Mowbray_thesis}.  Such wavelengths are comparable to the inter-atomic spacing where the hydrodynamic model is
likely to break down anyway, so we shall neglect such a contribution to the Hamiltonian in Eq.~(\ref{H}).

We further assume that the interaction with the external particle can be considered as a small perturbation. Hence, we can
expand the Hamiltonian in Eq.~(\ref{H}) to the second order with respect to the perturbation by declaring $\Vext =\lambda \Vext
$ and writing $\vb_\nu(\rbs,t)=\lambda\delta\dxdt_\nu(\rbs,t)+\bigoh(\lambda^2)$ and \(n_\nu(\rbs,t) = n_\nu^0 +
\lambda\deltan_\nu(\rbs,t)+\bigoh(\lambda^2)\), where $n_\nu^0$ is the ground-state electron density in the $\nu$th fluid. We
can further express the perturbed electron density in terms of the displacement field in a manner that will automatically
satisfy the linearized continuity equation by writing $\deltan_\nu(\rbs,t)=-n_\nu^0\grad\cdot\delta\xb_\nu(\rbs,t)$. Here the
gradient $\grad$ differentiates only in the directions tangential to the surface of the cylinder $r=R$. Moreover, by restricting
consideration to electrostatic phenomena, the velocity field will be irrotational. This will be ensured by defining a potential
function for the displacement field, $\xi_\nu(\rbs,t)$, such that $\delta\xb_\nu(\rbs,t)=-\grad\xi_\nu(\rbs,t)$. We can now
write the second-order Hamiltonian in terms of the function $\xi_\nu(\rbs,t)$ alone,
\begin{widetext}
  \begin{eqnarray}
    \label{eq:2}
\Hop_2&=& \sum_{\nu}n^0_\nu \int\!d^2\rbs \, \left\{ \frac{m_\nu^*}{2}
  \left( \|\grad\dotxi_\nu\|^2+\omega_{\nu
r}^2\|\grad\xi_\nu\|^2 \right) +\Vext(\rbs,t)\grad^2\xi_\nu \right\}
\nonumber\\
&+&  \sum_{\nu} \int\!d^2\rbs\, \left( \frac{\pi \hslash^2}{2m_\nu^*} -
  \frac{\charge^2}{\sqrt{2\pi n^0_\nu}} \right)
\left(n^0_\nu\grad^2\xi_\nu \right)^2
\nonumber\\
&+& \frac{\charge^2}{2} \sum_{\nu,\nu'} \iint\!d^2\rbs\,d^2\rbs'\, \frac{n^0_\nu \,n^0_{\nu'}}{\|\rbs -\rbs'\|} \grad^2\xi_\nu
\grad^2\xi_{\nu'}. \qquad \label{H2}
\end{eqnarray}
\end{widetext}

Before proceeding, it is worth mentioning that, while quantum treatment generally requires a second-order Hamiltonian
\cite{Lundqvist}, as in Eq.\ (\ref{eq:2}), such an approximation for the Hamiltonian yields semi-classical equations of motion
for the electron gas that correspond to the linearized hydrodynamic model. To illustrate this point, we note that one can derive
from $\Hop_2$ a Lagrange's equation of motion for the potential function $\xi_\nu(\rbs,t)$ as follows
\begin{eqnarray}
 \label{eq:3}
\ddotxi_\nu+\omega_{\nu
  r}^2\xi_\nu-s_\nu^2\grad^2\xi_\nu=\frac{1}{m_\nu^*}\left[\Vext(\rbs,t)-\charge\Phi_\text{ind}(\rbs,t)\right],
\end{eqnarray}
where we have defined the speed of propagation of density disturbances in the $\nu$th fluid, $s_\nu$, by
\begin{eqnarray}
\label{eq:4} s_\nu^2=\frac{2}{m_\nu^*}\left( \frac{\pi \hslash^2}{2m_\nu^*} n^0_\nu - \charge^2\sqrt{ \frac{n^0_\nu}{2\pi} }
\right).
\end{eqnarray}
Here $\Phi_\text{ind}(\rb,t)$ is the induced potential in the system due to polarization of the electron fluids, satisfying the
Poisson equation in three dimensions,
\begin{eqnarray}
\label{eq:5} \nabla^2\Phi_\text{ind}(\rb,t)=4\pi e \delta(r-R)\sum_{\nu=\pi,\sigma}n^0_\nu\grad^2\xi_\nu(\rbs,t),
\end{eqnarray}
where we have used the fact that the induced density in the $\nu$th fluid is given by $\deltan_\nu=n^0_\nu\grad^2\xi_\nu$ to the
first order. It is interesting to note that the effect of the Dirac term in Eq.~(\ref{eq:4}) is to reduce the adiabatic limit of
the speed due to the TF term, which can be written as $s_{TF\nu}=v_{F\nu}/\sqrt{2}$ with $v_{F\nu}=(\hslash/m_\nu^*)\sqrt{2\pi
n^0_\nu}$ being the Fermi speed for the $\nu$-th fluid in equilibrium. It can be shown that, in a hydrodynamic model without the
Dirac correction \cite{Stockli_2001,Wang_2004,Mowbray_2004}, the high-frequency expression for the speed due to the TF
interaction should be corrected so that $s_{TF\nu}=(\sqrt{3}/2)v_{F\nu}$ \cite{Atwal_2002}. However, including this effect in
the present model would require using a relaxation approximation for the kinetics of the electron gas, rendering second
quantization of the model unsuitable \cite{Atwal_2002}.

We make further progress by seeking a Fourier series representation of the potential function
$\xi_\nu(\rbs,t)\equiv\xi_\nu(\varphi,z,t)$ for its dependence on coordinates on the nanotube's surface. This will give rise to
a change in variables $\{\varphi,z\} \mapsto \{m,k\}$ for all quantities of interest, with $m$ and $k$ defining the modes of
plasmon propagation around the nanotube's circumference and in its axial direction, respectively. We initially assume that the
nanotube has a finite length $L$, so that its total area is $\Area=2\pi RL$ and $k$ is a discrete variable, taking values
$k=\frac{2\pi}{L}\ell$ with $\ell$ being an integer. Thus, defining the Fourier coefficients as
\begin{eqnarray}
\label{eq:6} \xfb_\nu(m,k,t)=R\int_{-\pi}^{\pi}d\varphi\int_{-L/2}^{L/2}dz\,\exponent^{-im\varphi-ikz}\xi_\nu(\varphi,z,t)
\equiv\int\!d^2\rbs\,\exponent^{-im\varphi-ikz}\xi_\nu(\rbs,t),
\end{eqnarray}
we may write
\begin{eqnarray}
\label{eq:7} \xi_\nu(\rbs,t)=\frac{1}{\Area}\sum_{m=-\infty}^{\infty}\sum_{k}\exponent^{im\varphi+ikz}\xfb_\nu(m,k,t)
\rightarrow \frac{1}{2\pi
R}\sum_{m=-\infty}^{\infty}\int_{-\infty}^{\infty}\frac{dk}{2\pi}\exponent^{im\varphi+ikz}\xfb_\nu(m,k,t),
\end{eqnarray}
where the rightmost expression involves a Fourier transform in the axial direction that arises in the limit
$L\rightarrow\infty$, when $k$ becomes a continuous variable. We note that, while the present formalism is most clearly
developed by using summation over $k$ for a nanotube of finite length, our final results will be computed by integration over
$k$, corresponding to the limit of an infinitely long nanotube. In this way we neglect any end effects on plasmon excitation
spectra, which may be justified by invoking a typically high aspect ratio for CNTs, $L/R\gg 1$.

One can now express $\Hop_2$ in Eq.~(\ref{H2}) as
\begin{eqnarray}
\label{eq:8}
\Hop_2&=&\frac{1}{\mathcal{A}}\sum_{m,k}\left(\frac{m^2}{R^2}+k^2\right)\left[h_{mk}^{(0)}+h_{mk}^{(\mathrm{int})}\right],
\end{eqnarray}
with the unperturbed part of the Hamiltonian given by
\begin{eqnarray}
\label{eq:9} h_{mk}^{(0)}&=&\sum_{\nu}n_\nu^0 \frac{m_\nu^*}{2} \left[ \vert{\bf \dot{\xfb}}_\nu\vert^2+\omega_{\nu
r}^2\vert\xfb_\nu\vert^2 +s_\nu^2\left(\frac{m^2}{R^2}+k^2\right)\vert\xfb_\nu\vert^2 \right]
\\
\nonumber
 &+&\frac{\charge^2}{2}Rg_{mk}\left(\frac{m^2}{R^2}+k^2\right)\sum_{\nu,\nu'}n_\nu^0n_{\nu'}^0\xfb_\nu^*\xfb_{\nu'},
\end{eqnarray}
and the interacting part by
\begin{eqnarray}
\label{eq:10} h_{mk}^{(\mathrm{int})}=-\sum_{\nu}n_\nu^0\Re\{\xfb_\nu^*\widetilde{V}_{\mathrm{ext}}\}, \label{hint}
\end{eqnarray}
where we have suppressed the dependencies on $(m,k,t)$ in $\xfb_\nu$ and its complex conjugate $\xfb_\nu^*$. Here, $g_{mk}\equiv
4\pi I_m(|k|R) K_m(|k|R)$ is the Green's function for the Poisson equation in cylindrical coordinates evaluated at the
nanotube's surface, with $I_m$ and $K_m$ being the modified Bessel functions of integer order $m$, of the first and second kind,
respectively. Moreover, $\widetilde{V}_{\mathrm{ext}}\equiv\widetilde{V}_{\mathrm{ext}}(R;m,k,t)$ is the Fourier transform with
respect to coordinates of the external potential $\Vext(\rbs,t)\equiv\Vext(R;\varphi,z,t)$, evaluated on the nanotube's surface.
In particular, for a point charge $eZ$ moving on a trajectory $\rb_0(t) = \{r_0(t),\varphi_0(t),z_0(t)\}$ that remains external
to the nanotube at all times, $r_0(t)>R$, we obtain
\begin{equation}
  \label{eq:33}
  \widetilde{V}_{\rm ext} (R;m,k,t) = -\charge^2Z\,Rg_{mk}(R,r_0(t))
  \,\exponent^{-im\varphi_0(t)-ikz_0(t)},
\end{equation}
where $g_{mk}(R,r_0)=4\pi I_m(|k|R) K_m(|k|r_0)$.

We now proceed to diagonalize $h_{mk}^{(0)}$ in the subspace of interacting $\sigma$ and $\pi$ fluids for fixed $m$ and $k$.
(The procedure is similar to that followed by Gorokhov {\it et al.}~\cite{Gorokhov_1996} in their treatment of C$_{60}$ plasmon
excitation with a  quantized two-fluid model). To this end, we define the factors
\begin{equation}
\label{eq:11} f_\nu=\frac{n^0_\nu m_\nu^*}{n_0m_*}= \left\{
\begin{array}{ll}
\frac{3}{4}\frac{m_\sigma^*}{m_*} & {\rm for\,\, \sigma~electrons}  \\
[0.5cm] \frac{1}{4}\frac{m_\pi^*}{m_*} & {\rm for\,\, \pi~electrons},
\end{array}
\right.
\end{equation}
where $n_0=n^0_\sigma + n^0_\pi$ is the total areal density of valence electrons in graphene, with relative weights $\nicefrac{3
}{4}$ and $\nicefrac{1 }{4}$ corresponding to the $sp^2$ hybridization \cite{Calliari}, and $m_*$ is a suitably defined mean
effective electron mass (see below). Thus, we can rewrite Eq.~(\ref{eq:9}) as
\begin{eqnarray}
\label{eq:12} h_{mk}^{(0)}=n_0\frac{m_*}{2}\left[f_\sigma\left(\vert{\bf
\dot{\xfb}}_\sigma\vert^2+\omega_{\sigma}^2\vert\xfb_\sigma\vert^2\right) + f_\pi\left(\vert{\bf
\dot{\xfb}}_\pi\vert^2+\omega_{\pi}^2\vert\xfb_\pi\vert^2\right)+\Delta\sqrt{f_\sigma
f_\pi}\left(\xfb_\sigma^*\xfb_\pi+\xfb_\pi^*\xfb_\sigma\right)\right].
\end{eqnarray}
where we have defined plasma frequencies of non-interacting fluids, $\omega_\sigma$ and $\omega_\pi$, by
\begin{equation}
\label{eq:13}
  \omega^2_{\nu}=\omega_{\nu r}^2 +
   \left(s_\nu^2+\charge^2Rg_{mk}\frac{n_\nu^0}{m_\nu^*}\right)\left(\frac{m^2}{R^2}+k^2\right),
\end{equation}
while their coupling is defined by
\begin{equation}
\label{eq:14}
   \Delta^2 = \charge^2Rg_{mk}\sqrt{\frac{n_\sigma^0}{m_\sigma^*}\frac{n_\pi^0}{m_\pi^*}}\left(\frac{m^2}{R^2}+k^2\right).
\end{equation}

One can diagonalize the Hamiltonian $h^{(0)}_{mk}$ by substituting
\begin{equation}
\label{eq:15}
  \sqrt{f_\sigma}\,{\widetilde{\xi}_\sigma} = A_1\cos\alpha -A_2\sin\alpha,
\end{equation}
\begin{equation}
\label{eq:16}
  \sqrt{f_\pi}\,{\widetilde{\xi}_\pi} =
  A_1\sin\alpha +A_2\cos\alpha,
\end{equation}
into Eq.~(\ref{eq:12}) and by choosing the angle $\alpha$ so that oscillations with amplitudes $A_1$ and $A_2$ are decoupled.
This is achieved when $\cot(2\alpha) = \frac{\omega_\sigma^2-\omega_\pi^2}{2\Delta^2}$, so that the non-interacting part of the
Hamiltonian can be written as a sum of decoupled oscillators,
\begin{equation}
  \label{eq:17}
h_{mk}^{(0)}=n_0\frac{m_*}{2}\sum_{j=1}^2\left[\dot{A}^*_{jmk}(t)\dot{A}_{jmk}(t) + \omega_{jmk}^2
  A_{jmk}^*(t)A_{jmk}(t) \right],
\end{equation}
where we have restored the dependencies on $(m,k,t)$ and defined the eigen-frequencies, $\omega_{jmk}>0$, of the decoupled
oscillators by
\begin{equation}
  \label{eq:18}
  \omega^2_{jmk} = \frac{\omega_\sigma^2+\omega_\pi^2}{2} \pm
  \sqrt{\left(\frac{\omega_\sigma^2-\omega_\pi^2}{2}\right)^2+\Delta^4},
\end{equation}
with $j=1,2$ for the positive and negative signs, respectively.

For the interacting Hamiltonian, we substitute Eqs.~(\ref{eq:15}) and (\ref{eq:16}) into Eq.~(\ref{hint}) and obtain
\begin{eqnarray}
 \label{eq:19}
  h_{mk}^{\rm (int)}& = & -
  \frac{1}{2}\frac{n_\sigma^0}{\sqrt{f_\sigma}}\left[\left(A_1\cos\alpha- A_2\sin\alpha \right)\widetilde{V}_{\rm ext}^* + c.c.\right]
  \nonumber \\
  &-&  \frac{1}{2}\frac{n_\pi^0}{\sqrt{f_\pi}}\left[\left(A_1\sin\alpha + A_2\cos\alpha\right) \widetilde{V}_{\rm ext}^* +
    c.c. \right],
\end{eqnarray}
which can be written in a more compact form if we define $m_*$ to be a weighted harmonic mean of the effective masses in the
$\sigma$ and $\pi$ fluids,
\begin{equation}
\label{eq:20} \frac{1}{m_*}=\frac{1}{n_0}\left(\frac{n_\sigma^0}{m_\sigma^*}+\frac{n_\pi^0}{m_\pi^*}\right),
\end{equation}
as follows
\begin{eqnarray}
\label{eq:21} h_{mk}^{\rm (int)}& = & \frac{n_0}{2}\left\{\left[D_{1mk}A_{1mk}(t)+D_{2mk}A_{2mk}(t)\right] \widetilde{V}_{\rm
ext}^*(R;m,k,t)+c.c.\right\},
\end{eqnarray}
where we have restored the dependencies on $(m,k,t)$. Here
\begin{eqnarray}
\label{eq:22} D_{jmk} = \left\{
  \begin{array}{cc}
    -\cos(\alpha-\beta), & \,\,j=1, \\
    \sin(\alpha-\beta), & \,\,j=2,
    \end{array}
  \right.
\end{eqnarray}
where the angles $\alpha$ and $\beta$ are defined, respectively, by
\begin{equation}
 \label{eq:24}
  \left\{\begin{array}{c}
\cos\alpha \\
\sin\alpha
\end{array} \right\} =
\frac{1}{\sqrt{2}} \sqrt{1\pm\frac{\omega_\sigma^2-\omega_\pi^2}{\omega_{1mk}^2-\omega_{2mk}^2}}
\end{equation}
and
\begin{eqnarray}
\label{eq:23} \cos\beta=\sqrt{\frac{n_\sigma^0m_*}{n_0m_\sigma^*}},\quad\quad\sin\beta=\sqrt{\frac{n_\pi^0m_*}{n_0m_\pi^*}}.
\end{eqnarray}

It is worth mentioning that a classical, single-fluid model is recovered by letting both $\omega_{\nu r}\to 0$ and $s_\nu\to 0$
in Eq.\ (\ref{eq:13}), so that Eq.\ (\ref{eq:18}) then gives $\alpha=\beta$ with $\omega_{2mk}=0$ and $\omega_{1mk}$
corresponding to plasma frequency of a 2D electron gas with surface density $n_0=n_\sigma^0+n_\pi^0$ and an effective mass $m_*$
defined via Eq.\ (\ref{eq:20}) \cite{Wang_2004,Mowbray_2004}.

\subsection{Quantization}

Once we have the decoupled Hamiltonian to describe the system, it is useful to apply a quantum treatment in terms of creation
and annihilation operators. This provides a simple and clear  way to describe the excitation of  plasmons (oscillators) due to
the interaction with the  external particle. Following the quantization procedure described by Arista and Fuentes
\cite{Arista_2001}, we assign to the coefficients $A_{jmk}$ creation and annihilation operators as follows
\begin{equation}
\label{quant}
 A_{jmk}(t) \longrightarrow
  \frac{\gamma_{jmk}}{2\omega_{jmk}}\left[\hat{a}^\dagger_{jmk}(t) + \hat{a}_{jmk}(t)\right],
\end{equation}
where the coefficient $\gamma_{jmk}$ is to be determined. The operators $\hat{a}_{jmk}$ and $\hat{a}^\dagger_{jmk}$ satisfy the
usual commutation relations
$$
\left[\hat{a}_{j'm'k'},\hat{a}^\dagger_{jmk}\right] =
  \delta_{jj'}\delta_{kk'}\delta_{mm'},
$$
while their time dependence, $\hat{a}_{jmk}(t) = \hat{a}_{jmk}\exp\left( -i\omega_{jmk}t\right),$ gives $\hat{\dot{a}}_{jmk}(t)
= -i\omega_{jmk}\hat{a}_{jmk}(t)$, so that
\begin{equation}
\label{eq:26} \dot{A}_{jmk}(t) \longrightarrow
  i\frac{\gamma_{jmk}}{2}\left[\hat{a}^\dagger_{jmk}(t) -
    \hat{a}_{jmk}(t)\right].
\end{equation}

With these relations, the (quantized) non-interacting Hamiltonian of the mode $(m,k)$ can be written as
\begin{eqnarray}
  \label{eq:27}
  \hat{h}^{(0)}_{mk}
=n_0\frac{m_*}{2}\sum_{j=1}^2
 \gamma_{jmk}^2
 \left(\hat{a}^\dagger_{jmk}\hat{a}_{jmk}+\frac{1}{2}\right)
\end{eqnarray}
giving for the full non-interacting Hamiltonian (see Eq. (9))
\begin{eqnarray}
  \label{eq:28}
  \hat{\Hop}^{(0)}_2 &=&
  \frac{1}{\mathcal{A}} \sum_{m,k}
 \left(\frac{m^2}{R^2} + k^2\right) \hat{h}_{mk}^{(0)} \nonumber \\
  &=& \sum_{j,m,k} \hslash\omega_{jmk}
  \left(\hat{a}^\dagger_{jmk}\hat{a}_{jmk}+\frac{1}{2}\right).
\end{eqnarray}
This procedure gives the coefficient $\gamma_{jmk}$ to be used in Eq.~(\ref{quant}) as
\begin{equation}
  \label{eq:29}
  \gamma_{jmk} =
  \sqrt{\frac{2\mathcal{A}\hslash\omega_{jmk}}{m_* n_0
      \left(\frac{m^2}{R^2}+k^2\right)}}.
\end{equation}

For the (quantized) interacting Hamiltonian of the mode $(m,k)$ we obtain
\begin{eqnarray}
\label{eq:30} \hat{h}_{mk}^{\rm (int)} = \frac{n_0}{2}\sum_{j=1}^2\frac{\gamma_{jmk}}{2\omega_{jmk}}D_{jmk}\left\{ \left[
\hat{a}^\dagger_{jmk}(t) + \hat{a}_{jmk}(t) \right] \widetilde{V}_{\rm ext}^*(R;m,k,t)+h.c. \right\},
\end{eqnarray}
so that the total interacting Hamiltonian reads
\begin{equation}
  \label{eq:31}
  \hat{\Hop}_2^{\rm (int)} = \sum_{j,m,k}
  \Gamma_{jmk}(t)\left[\hat{a}^\dagger_{jmk} (t) + \hat{a}_{jmk}(t) \right],
 \end{equation}
with
\begin{equation}
  \label{eq:32}
  \Gamma_{jmk}(t) = \sqrt{
    \frac{\hslash
      n_0\left(\frac{m^2}{R^2}+k^2\right)}{2m_*\mathcal{A}\omega_{jmk}}} \, D_{jmk} \Re\left[\widetilde{V}_{\rm ext} (R;m,k,t)\right].
\end{equation}
We note that the interacting Hamiltonian in Eq.\ (\ref{eq:31}) is given by a superposition of the linear displacements of
plasmons as quantum oscillators, which is a consequence of using the second-order approximation for the full Hamiltonian, given
in Eq.\ (\ref{eq:2}) \cite{Lundqvist}.

We further follow the formalism presented by Arista and Fuentes \cite{Arista_2001} and obtain the average number of excited
plasmons in a given mode $(j,m,k)$ as
\begin{equation}
  \label{eq:34}
\mu_{jmk}=\left|\frac{1}{\hslash}\int_{t_0}^t
    dt'\,\exponent^{i\omega_{jmk}t'}\Gamma_{jmk}(t')\right|^2,
\end{equation}
where $t_0\rightarrow-\infty$ and $t\rightarrow+\infty$. Inserting Eq.~(\ref{eq:32}) in Eq.~(\ref{eq:34}), we obtain
\begin{equation}
  \label{eq:35}
\mu_{jmk} = C_{jmk}\,\left|\VV\left(R;m,k,\omega_{jmk}\right)\right|^2,
\end{equation}
where
\begin{equation}
  \label{eq:36}
  C_{jmk} =
  \frac{n_0}{\mathcal{A}}\,\frac{D_{jmk}^2}{2\hslash m_*}
  \,\frac{\frac{m^2}{R^2}+k^2}{\omega_{jmk}},
\end{equation}
and $\VV(R;m,k,\omega)$ is the Fourier transform with respect to time of the function $\widetilde{V}_{\rm ext}(R;m,k,t)$,
\begin{equation}
  \label{eq:37a}
  \VV(R;m,k,\omega)=\int_{-\infty}^\infty dt \, \exponent^{i\omega t}
  \widetilde{V}_{\rm ext}(R;m,k,t).
\end{equation}

The time integral in Eq.~(\ref{eq:37a}) can be solved analytically by using Eq.~(\ref{eq:33}) for the case of a straight-line
trajectory with constant velocity, given by $\rb_0(t) = \{r_0(t),\varphi_0(t),z_0(t)\}$ with
\begin{eqnarray}
\nonumber
r_0(t) &=& \sqrt{r_{\rm min}^2 + v_\perp^2 t^2},\\
\nonumber
\varphi_0(t) &=& \arctan\left(\frac{v_\perp t}{r_{\rm min}}\right),\\
\nonumber z_0(t) &=& v_\| t,
\nonumber
\end{eqnarray}
where $r_{\rm min}>R$ is the shortest distance between the trajectory and the nanotube axis, and $v_\bot$ and $v_\|$ are the
components of the incident particle's velocity in the directions perpendicular and parallel to the nanotube's axis,
respectively. Thus, one obtains \cite{Mowbray_thesis}
\begin{equation}
  \label{eq:37}
\VV(R;m,k,\omega)=-
    4\pi \charge^2ZR
    I_m(|k|R) \mathcal{K}_m(k,\omega),
\end{equation}
where
\begin{equation}
  \label{eq:38}
\mathcal{K}_m(k,\omega) = \pi\,\frac{\exponent^{-\frac{r_{\rm
min}}{v_\bot}\sqrt{(\omega-kv_\|)^2+(kv_\bot)^2}}}{\sqrt{(\omega-kv_\|)^2+(kv_\bot)^2}}
\left(\frac{\omega-kv_\|+\sqrt{(\omega-kv_\|)^2+(kv_\bot)^2}}{\vert k\vert v_\bot}\right)^m.
\end{equation}
Here, the velocity components may be expressed in terms of the total projectile speed, $v$, and the incident angle, $\theta$,
relative to the nanotube axis as $v_\bot=v\sin\theta$ and $v_\|=v\cos\theta$. Without any loss of generality, one may adopt the
range of angles $0\le\theta\le\pi/2$, so that the direction of projectile motion is defined by positive values $v_\bot>0$ and
$v_\|>0$, corresponding to the directions of increasing variables $\varphi$ and $z$, respectively. As a consequence, one can
then infer from Eq.~(\ref{eq:38}) that the propagation direction of plasmon modes $(m,k)$, which is commensurate with the
direction of projectile motion, will be given by positive values, $m>0$ and $k>0$.

Finally, using the fact that the plasma excitations by an external particle are represented by independent quantum oscillators
of the mode $(j,m,k)$, and that the probability distribution of exciting such an oscillator to the $N$th state is Poissonian
defined by the mean $\mu_{jmk}$, one can deduce several statistical properties of practical interest. For example, it is shown
in appendix~\ref{app:a} that the probability distribution of plasma excitations in the mode with fixed $j$ and $m$ for arbitrary
$k$ is also Poissonian, defined by the mean
\begin{equation}
  \label{eq:39}
  \mu_{jm} = \sum_k\mu_{jmk}\rightarrow \frac{L}{2\pi}\int_{-\infty}^\infty dk\,\mu_{jmk}.
\end{equation}
Moreover, we shall derive in appendix~\ref{app:a} an expression for the probability density for energy loss, $P(\vareps)$ in
Eq.\ (\ref{eq:b17}), as well as the total energy loss, $\Eloss$ in Eq.\ (\ref{eq:b10}), which can be compared with experiments
or other theoretical approaches.

A few comments may be in order about the projectile trajectory in relation to plasmon excitations of the nanotube. The
straight-line trajectory presents a good approximation for a projectile with sufficiently high momentum, so that the effects of
the recoil and trajectory bending may be neglected. On the other hand, a projectile moving at very high speeds would require
inclusion of the retardation effects in the electron response of the CNT, which is beyond the scope of the present work.
Moreover, our neglect of the end effects on plasmon excitation may be justified when the projectile moves on non-parallel
trajectories with angles $\tan\theta \gg R/L$. In practice, this condition is satisfied for angles $\theta\gtrsim 1^\circ$ owing
to the high aspect ratio of CNTs.

The case of parallel trajectory with $\theta=0$ deserves special attention. In that case, one has to assume that the nanotube
length $L$ is finite, but still large enough to allow the neglect of end effects on plasmon excitation, as indicated in the
limiting form of Eq.\ (\ref{eq:7}). On the other hand, it is rigorously shown in the last paragraph of appendix~\ref{app:b}
that, when $v_\bot=0$, the number of excited plasmons in the mode $(j,m,k)$ becomes proportional to
$L\,\delta\!\left(\omega_{jmk}-kv_\|\right)$, and Eq.\ (\ref{eq:35}) must be then evaluated by using Eq.\ (\ref{eq:c16}), rather
than Eqs.\ (\ref{eq:37}) and (\ref{eq:38}). If $L$ may also be considered short enough, one can further define the stopping
force, or the energy loss per unit path length, $S=\Eloss/L$, giving Eq.\ (\ref{eq:c17}) for a projectile moving on a parallel
trajectory outside the nanotube. That result is extended in a straightforward manner in Eq.\ (\ref{eq:c18}) for a particle
channeled inside the nanotube \cite{Zhou_2006,Borka_2006}. In either case, the quantity $S$ may be considered independent of $L$
if the total energy loss $\Eloss$ in a finite-length nanotube is much smaller than the initial kinetic energy of the particle on
a parallel trajectory.

\section{Results and Discussion}\label{III}
\begin{figure}
\includegraphics[width=0.5\textwidth]{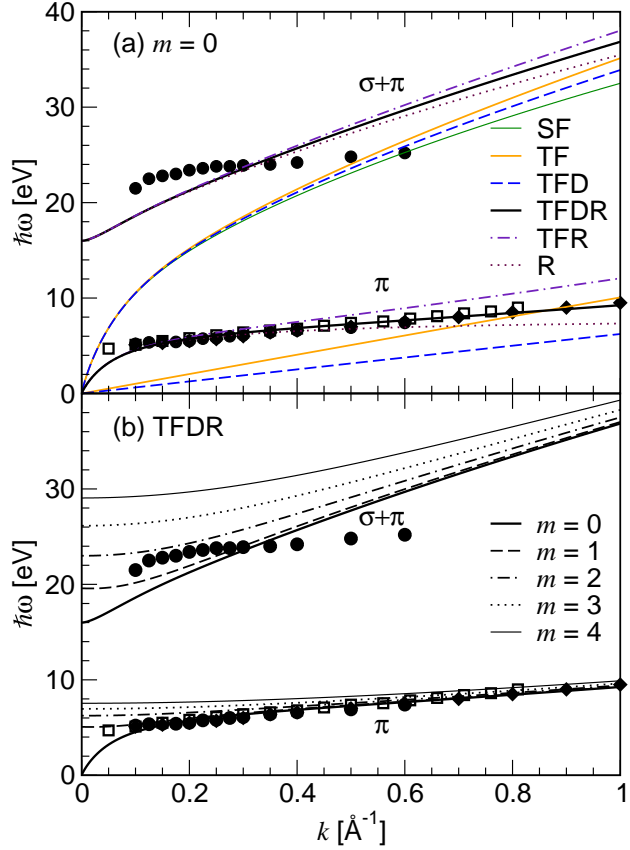}
\caption{Plasmon energies \(\hslash\omega\) in eV versus longitudinal
  momentum \(k\) in \AA$^{-\text{1}}$ of a SWCNT with radius
  $R=7$~\AA\ for (a) angular momentum \(m = \text{0}\) from
  single-fluid (SF)  \cite{Wang_2004}, two-fluid Thomas-Fermi (TF)
  \cite{Cazaux_1970,Mowbray_2006}, two-fluid Thomas-Fermi-Dirac (TFD),
  restoring frequency (R)~\cite{Barton_1993}, two-fluid Thomas-Fermi
  with restoring frequency (TFR), and two fluid Thomas-Fermi-Dirac
  with restoring frequency (TFDR) models, and (b) two fluid
  Thomas-Fermi-Dirac with restoring frequency model for angular
  momenta \(m = \) 0, 1, 2, 3, and 4.  Experimental results from
Ref.~\cite{Pichler_1998} (circles), Ref.~\cite{Kramberger_2008} (diamonds), and Ref.~\cite{Kramberger_2008b} (squares) are
provided for comparison. } \label{fig:1}
\end{figure}

\begin{figure}
\includegraphics[width=0.9\textwidth]{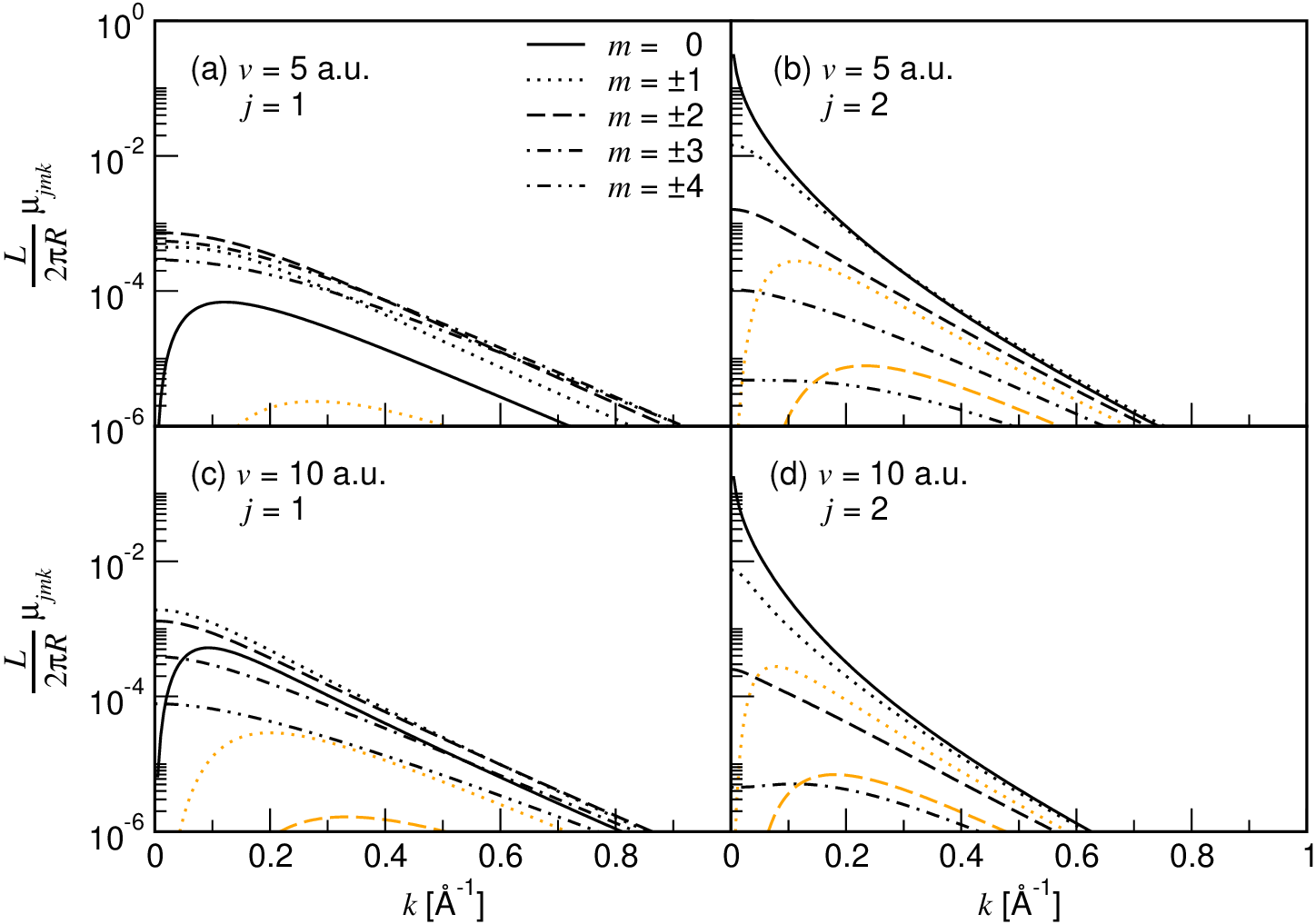}
\vspace*{2mm} \caption{Average number of excited plasmons $\frac{L}{2\pi R}\mu_{jmk}$ as a function of longitudinal momentum $k$
in \AA$^{-\text{1}}$ on a SWCNT with $R = \text{7}$ \AA\ for (a,c) $\sigma+\pi$ modes ($j = \text{1}$) and (b,d) $\pi$ modes ($j
= \text{2}$) with angular momenta $|m| =$ 0 (\textbf{---$\!$---}), 1 ($\cdots\cdots$), 2   (-- -- --),  3 (-- $\cdot$ --), and 4
(-- $\cdot\cdot$ --), shown by black lines when $m\ge 0$ and orange lines when $m<0$, due to a proton with $r_{\min{}} =
\text{10.5}$ \AA, incident at angle relative to the SWCNT axis of  $\theta = \text{90}^{\circ}$ with speed (a,b) $v = \text{5}$
a.u. and (c,d) $v = \text{10}$ a.u.}  \label{fig:2}
\end{figure}

\begin{figure}
\includegraphics[width=0.9\textwidth]{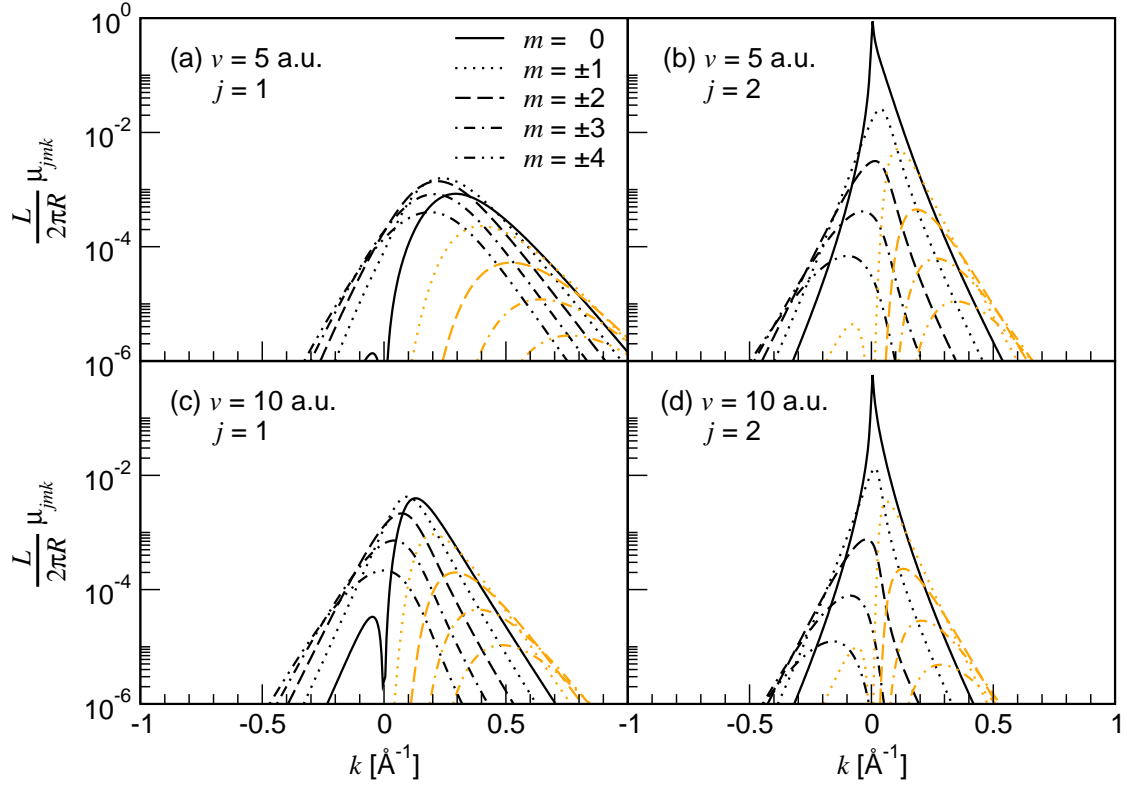}
\vspace*{2mm} \caption{Average number of excited plasmons $\frac{L}{2\pi R}\mu_{jmk}$ as a function of longitudinal momentum $k$
in \AA$^{-\text{1}}$ on a SWCNT with $R = \text{7}$ \AA\ for (a,c) $\sigma+\pi$ modes ($j = \text{1}$) and (b,d) $\pi$ modes ($j
= \text{2}$) with angular momenta $|m| =$ 0 (\textbf{---$\!$---}), 1 ($\cdots\cdots$), 2   (-- -- --),  3 (-- $\cdot$ --), and 4
(-- $\cdot\cdot$ --), shown by black lines when $m\ge 0$ and orange lines when $m<0$, due to a proton with $r_{\min{}} =
\text{10.5}$ \AA, incident at angle relative to the SWCNT axis of  $\theta = \text{45}^{\circ}$ with speed (a,b) $v = \text{5}$
a.u. and (c,d) $v = \text{10 }$ a.u.}  \label{fig:3}
\end{figure}

\begin{figure}
\includegraphics[width=0.8\textwidth]{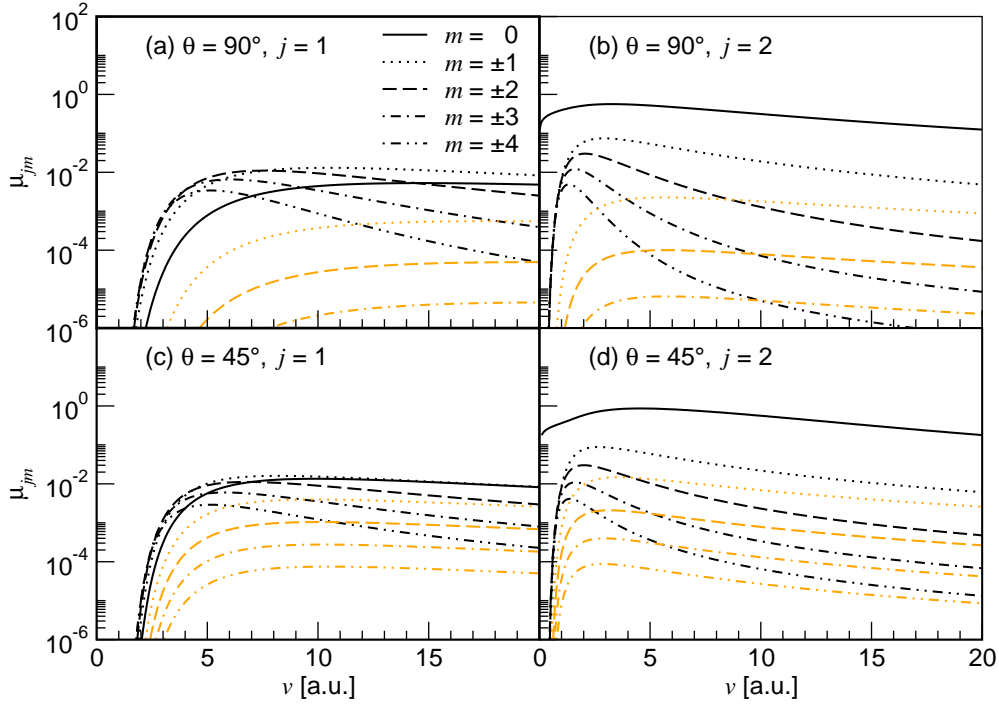}
\vspace*{2mm} \caption{ Average number of excited plasmons $\mu_{jm}$, defined in Eq.\ (\ref{eq:39}), on a SWCNT with $R =
\text{7}$ \AA\ for (a,c) $\sigma+\pi$ modes ($j = \text{1}$) and $\pi$ modes ($j = \text{2}$) with angular momenta $|m|= 0$
(\textbf{---$\!$---}), 1  ($\cdots\cdots$), 2 (-- -- --), 3  (-- $\cdot$ --), and 4 (-- $\cdot\cdot$ --), shown by black lines
when $m\ge 0$ and orange lines when $m<0$ as a function of the total speed $v$ in a.u.~of a proton with $r_{\min{}} =
\text{10.5}$ \AA\ at an angle relative to the SWCNT axis of (a,b) $\theta = \text{90}^{\circ}$ and (c,d) $\theta =
\text{45}^{\circ}$. } \label{fig:4}
\end{figure}

\begin{figure}
\includegraphics[width=0.8\textwidth]{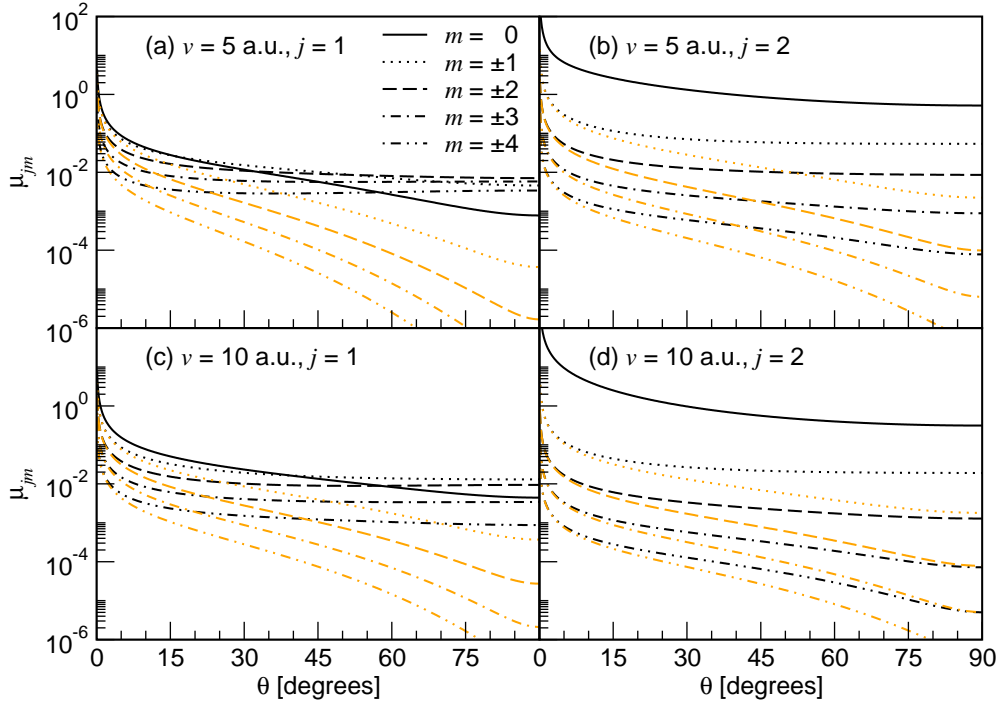}
\vspace*{2mm} \caption{ Average number of excited plasmons $\mu_{jm}$, defined in Eq.\ (\ref{eq:39}), on a SWCNT with $R =
\text{7}$ \AA\ for (a,c) $\sigma+\pi$ modes ($j = \text{1}$) and $\pi$ modes ($j = \text{2}$) with angular momenta $|m|= 0$
(\textbf{---$\!$---}), 1  ($\cdots\cdots$), 2 (-- -- --), 3  (-- $\cdot$ --), and 4 (-- $\cdot\cdot$ --), shown by black lines
when $m\ge 0$ and orange lines when $m<0$ as a function of the angle relative to the SWCNT axis $\theta$ in degrees for a proton
with $r_{\min{}} = \text{10.5}$ \AA\ at the total speed of (a,b) $v = \text{5}$ a.u.~and (c,d) $v = \text{10}$ a.u.}
\label{fig:5}
\end{figure}

\begin{figure}
\includegraphics[width=0.5\textwidth]{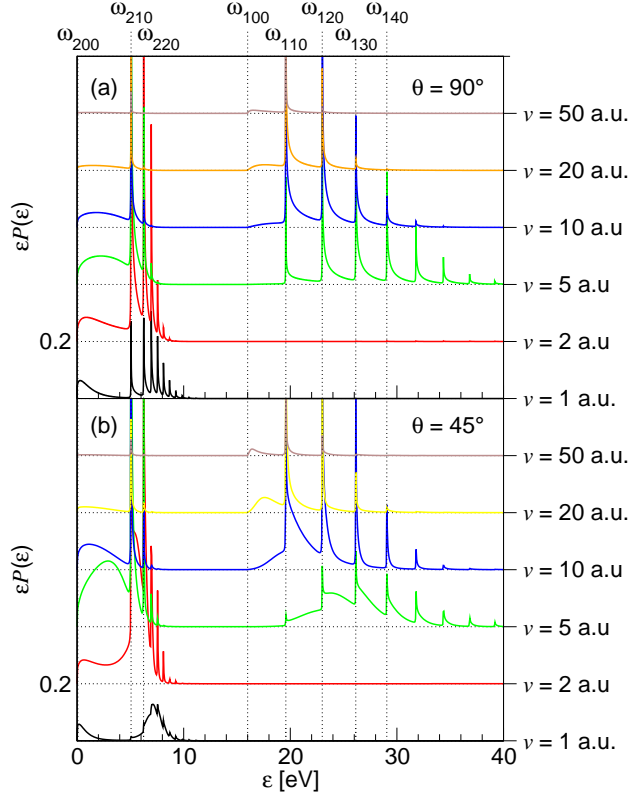}
\vspace*{2mm} \caption{Probability density for proton energy loss $\vareps$ in eV, expressed as \(\vareps P(\vareps)\), for
proton trajectory passing near a SWCNT with $R =\text{7}$ \AA\ at the minimum separation of $r_{\min{}} = \text{10.5}$~\AA\ with
an angle relative to the SWCNT axis of (a) $\theta = \text{90}^{\circ}$ and (b) $\theta = \text{45}^{\circ}$, at the total speed
of $v =$ 1, 2, 5, 10, 20, and 50 a.u.. The spectra are vertically shifted (in steps of 0.2) for the sake of clarity. The
position of the $k\rightarrow 0$ plasmon modes \(\omega_{jm0}\) are provided for comparison.} \label{fig:6}
\end{figure}

\begin{figure}
\includegraphics[width=0.5\textwidth]{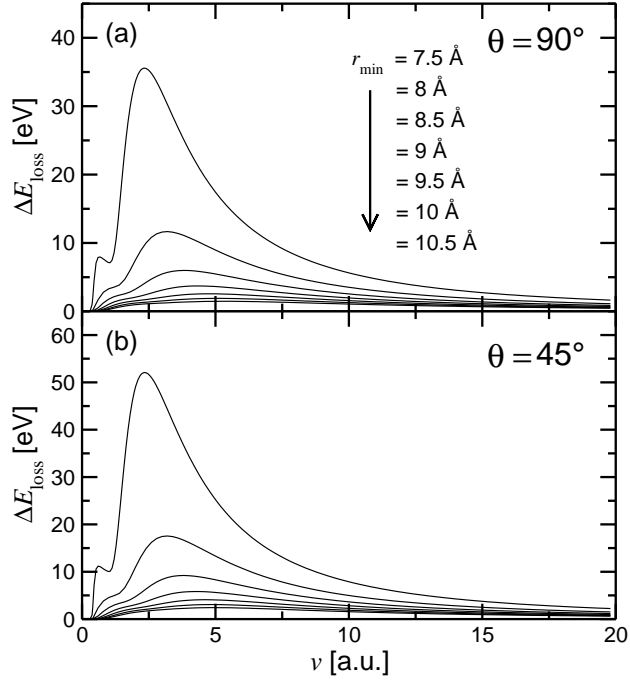}
\caption{Total energy loss \(\Delta E_{\text{loss}}\) in eV versus speed $v$ in a.u.~for proton trajectory passing near a SWCNT
with $R =\text{7}$ \AA\ with an angle relative to the SWCNT axis of (a) $\theta = \text{90}^{\circ}$ and (b) $\theta =
\text{45}^{\circ}$, at the minimum separation of $r_{\min{}} = $ 7.5~\AA\ (top), 8.0~\AA, 8.5~\AA, 9.0~\AA, 9.5~\AA, 10.0~\AA,
and 10.5~\AA\ (bottom). } \label{fig:7}
\end{figure}

\begin{figure}
\includegraphics[width=0.5\textwidth]{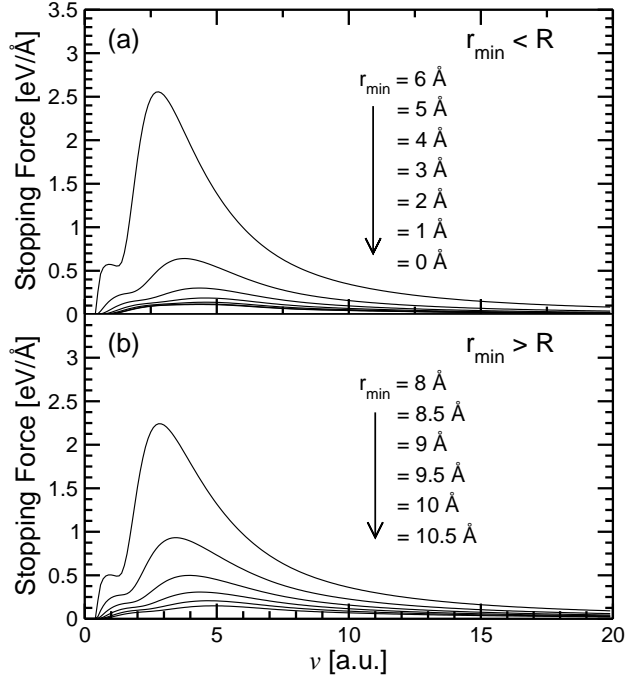}
\caption{Stopping force in eV/\AA\ versus speed $v$ in a.u.~for a proton travelling parallel to the axis of a SWCNT with $R =
\text{7}$ \AA\ (a) inside the SWCNT at $r_{\rm min} = 0$ (bottom), 0.5 \AA, 1.0 \AA, $\ldots$ , 6.0 \AA\ (top), and (b) outside
the SWCNT at $r_{\rm min} =$ 8.0 \AA\ (top), 8.5 \AA, \ldots, 10.5 \AA\ (bottom).} \label{fig:8}
\end{figure}

We first compare in Fig.~\ref{fig:1}(a) the effects of restoring (R), Thomas-Fermi (TF) and Dirac (D) interactions in Eq.\
(\ref{eq:1}) on the dispersion relations $\omega_{jm}(k)$, defined in Eq.~(\ref{eq:18}), for modes $j=1,2$ with $m=0$ in a CNT
with radius $R=7$~\AA, assuming that the effective masses are all equal to the free electron mass. We note that setting all
these interactions to zero causes the two fluids to collapse into a single classical electron fluid (SF) with surface density
$n_0=4n_\mathrm{at}$, giving rise to the $\sigma+\pi$ electron plasma oscillations \cite{Wang_2004}. Introducing either the R
interaction or the TF (with or without the D) interaction terms causes splitting of the plasmon dispersions into the
high-frequency ($j=1$ or $\sigma+\pi$) branch, which is very close to the single-fluid branch, and the low frequency ($j=2$ or
$\pi$) branch. The main difference between the splitting of plasmon branches due to the R interaction as opposed to the TF(D)
interaction is that, in the latter case, both the upper and the lower plasmon dispersions $\omega_{j0k}$ vanish as $k\rightarrow
0$, while the $\pi$ branch is markedly acoustic. On the other hand, inclusion of the R interaction renders the $\pi$ plasmon
branch quasi-acoustic, with an almost constant slope for $k\gtrsim$ 0.1 \AA$^{-1}$, in close agreement with the experimental
data \cite{Pichler_1998,Kramberger_2008,Kramberger_2008b}. It is also seen in Fig.~\ref{fig:1}(a) that the inclusion of the D
interaction both lowers the upper branch to some extent and reduces the slope of the lower branch, as expected from Eq.
(\ref{eq:4}).

Since in the rest of this work we use a ``complete'' two-fluid hydrodynamic model, which includes restoring (R), Thomas-Fermi
(TF) and Dirac (D) interactions, we compare in Fig.~\ref{fig:1}(b) its modes with $m=0$, 1, 2, 3, and 4 for both the upper
($j=1$ or $\sigma+\pi$) and lower ($j=2$ or $\pi$) plasmon branches with the experimental data
\cite{Pichler_1998,Kramberger_2008,Kramberger_2008b}. This semi-quantitative agreement with experiment, particularly for the
narrow group of lower plasmon branches, indicates that the inclusion of both the R and D interactions improves the present,
complete hydrodynamic model in comparison to the previous two-fluid model, which only included the TF interaction
\cite{Mowbray_2004}.

In the remaining calculations we consider a particle with $Z=\pm 1$ (proton or electron), passing by a SWCNT of radius
$R=7$~\AA\, at a minimum distance $r_{\min{}} = \text{10.5}$ \AA\, (unless indicated otherwise).

Figures \ref{fig:2} and \ref{fig:3} show the average number of excited plasmons, $\mu_{jmk}$, as a function of $k$ for $|m|=0$,
1, 2, 3, and 4 with $j=1$ (a,c) and $j=2$ (b,d), for two total speeds, $v=5$~a.u.~(a,b) and $v=10$~a.u.~(c,d), and for two
incident angles, $\theta=90^\circ$ (Fig.~\ref{fig:2}) and $\theta=45^\circ$ (Fig.~\ref{fig:3}). Note that, for normal incidence,
$\mu_{jmk}$ is an even function of wavenumber, so only positive $k$ values are shown in Fig.~\ref{fig:2}. We observe that, for
low $|k|$ values, the low-frequency branches ($j=2$) dominate over the high-energy modes ($j=1$) in both Fig.~\ref{fig:2} and
Fig.~\ref{fig:3}, but this trend seems to be reversed for increasing values of $|k|$, increasing speed, and oblique incidence.
In particular, the singular behavior of the mode $j=2$, $m=0$ as $k\to 0$ in Figs.\ \ref{fig:2} and \ref{fig:3} stems from the
way how its dispersion relation vanishes in the long wavelength limit in Fig.\ \ref{fig:1}, which may be approximated by a
one-dimensional plasmon dispersion for the $\pi$ electron fluid, $\omega_{20k}\sim k\sqrt{4\pi R
e^2(n^0_\pi/m^*_\pi)\ln(1.123/kR)}$ for $k\lesssim$ 0.1 \AA$^{-1}$ \cite{Sato}. At oblique incidence the $\mu_{jmk}$ curves are
no longer even functions of $k$, and the observed dominance in Fig.~\ref{fig:3} of the values with $k>0$ indicates that the
preferential direction for propagation of the plasmon coincides with the direction of the longitudinal velocity component of the
projectile, $v_\parallel>0$. In addition, one notices in Fig.~\ref{fig:2} that the modes with $m>0$ (black lines) are generally
excited with much higher probabilities than the modes with $m<0$ (orange lines) indicating that, for normal incidence, plasmon
waves tend to propagate around the nanotube circumference in the direction of projectile motion. However, the situation is
reversed for oblique incidence, so that the modes with $m<0$ tend to dominate over those with $m>0$ for increasing $k>0$ values
in Fig.~\ref{fig:3}.

Figure~\ref{fig:4} shows the dependence of the average number of excited plasmons, $\mu_{jm}$, which is obtained after
integration of the curves in Figs.~\ref{fig:2} and \ref{fig:3} over $k$, on the total particle speed $v$ for $\theta=90^\circ$
(a,b) and $\theta=45^\circ$ (c,d), for the branches $j=1$ (a,c) and $j=2$ (b,d). One can say that, globally, the mode $j=2$,
$m=0$ dominates for any speed or direction of the incident particle, displayed in Fig.~\ref{fig:4}, but we also observe that the
relative contributions of the other modes change with increasing $v$. For higher speeds of the incident particle, the relative
importance of the modes with $j=1$ increases, especially for $m=0$ and $m=1$. Moreover, one notices that the modes with $m<0$
are generally suppressed compared to the modes with $m>0$ at lower speeds, but they tend to regain some importance as the speed
increases. We also notice that there exist velocity thresholds for the excitation of the different non-acoustic modes (all $j=1$
modes, and the $j=2$ modes with $m\neq 0$). We remark that finite values of the average number $\mu_{jm}$ for the mode $j=2$,
$m=0$, which extend down to low speeds in Fig.~\ref{fig:4}, are likely a consequence of the singular behavior of the
corresponding number $\mu_{jmk}$ as $k\to 0$ seen in Figs.\ \ref{fig:2} and \ref{fig:3}, but they are not in violation of the
conservation of energy for slowly moving projectiles, as documented by the total energy loss at low speeds in
Fig.~\ref{fig:7}~\footnote{In fact, our computations confirmed that there is an abrupt decrease of the average number of the
$j=2$, $m=0$ plasmons as $v\to 0$, which is not rendered in Fig.~\ref{fig:4}.}.

The dependence of $\mu_{jm}$ on the incidence angle $\theta$ is displayed in Fig.~\ref{fig:5} for the modes with $j=1$ (a,c) and
$j=2$ (b,d) for two total speeds: $v=5$~a.u.~(a,b) and $v=10$~a.u.~(c,d). In general, the relative importance of various modes
is similar to that seen in Fig.~\ref{fig:4} exhibiting dominance of the mode $j=2$, $m=0$. One notices that the modes with
$m<0$, which are suppressed at finite angles of incidence, become equal to the modes with $m>0$ in the limit of grazing
incidence, $\theta\to 0$, as expected from symmetry considerations. Furthermore, the overall increase in the number of all
excited modes with decreasing angle $\theta$ may be ascribed to increasing interaction time with an otherwise infinite nanotube,
which should become proportional to $\mathrm{cosec}\,\theta$ for angles $\theta$ smaller than several degrees. In particular, it
is interesting to note that, while in Figs.~\ref{fig:2}, \ref{fig:3} and \ref{fig:4} the average numbers of excited plasmon
modes are generally small, and even the most dominant mode $j=2$, $m=0$ hardly exceeds unity, the number of plasmons excited in
the mode $j=2$, $m=0$ is seen in Fig.~\ref{fig:5} to greatly exceed unity for, e.g., $\theta \lesssim 5^\circ$.

From the results of previous section it is possible to obtain a statistical description of energy loss of the incident particle
(see appendix~\ref{app:a}). One of the most useful quantities is the probability density, $P(\vareps)$, for losing a given
amount of energy, $\vareps$, which can be related to the energy loss spectra with peaks corresponding to the excitation of
various plasmon modes. In Fig.~\ref{fig:6} we show the product $\vareps  P(\vareps)$ for several incident speeds of a particle
travelling perpendicular (a) and with an inclination angle of $45^\circ$ (b) with respect to the tube axis. We can see two
distinct groups of peaks, indicating excitation of the two branches of plasmons, $\pi$ and $\sigma+\pi$, shown in
Fig.~\ref{fig:1}. At low speeds ($v\sim 1-2$~a.u.), only the low-energy plasmons are excited~\footnote{Notice that for the case
of electrons  a recoil correction  must be considered for low velocities ($v\sim 1-2$~a.u.).}, which is commensurate with the
demonstrated ability of the low-energy EELS technique to probe the $\pi$ plasmon excitation in graphene \cite{Liu_2008}. On the
other hand, as the incident speed increases, we notice in Fig.~\ref{fig:6} excitation of both $\sigma+\pi$ and $\pi$ plasmons,
as observed in the high-energy EELS experiments \cite{Kramberger_2008,Kramberger_2008b,Eberlein_2008}. Even though we do not
pursue here the problem of plasmon excitations at relativistic projectile speeds, one may infer from Fig.~\ref{fig:6} that the
excitation probabilities of both groups of plasmons decrease as the speed exceeds a value on the order of 20 a.u., with the
low-energy plasmon peaks being almost diminished and the high-energy peaks still visible at $v\sim$ 50 a.u..

Looking into details of the spectra in Fig.~\ref{fig:6}, one notices that, as the speed increases, there are changes not only in
the overall intensity of the peaks, but also in the relative weight of each mode. At lower speeds, modes $m=2, 3, \ldots$ give
large contributions in both branches, while at higher speeds their weights are suppressed. It is interesting to notice that the
mode $j=1$, $m=0$ only appears at the speeds $v\gtrsim$ 10 a.u.\ as a broad feature just above the restoring frequency of
$16$~eV of the $\sigma$ fluid. When the incidence is oblique, we observe broadening of all peaks, especially at low speeds,
which is due to an increase of the high-$k$ contributions to the excitation of various plasmon modes (\emph{c.f.}
Figs.~\ref{fig:2} and \ref{fig:3}).

By integrating the curves shown in Fig.~\ref{fig:6}, one can obtain the total energy loss, $\Delta E_{loss}$, suffered by the
incident particle. This is shown in Fig.~\ref{fig:7} versus the particle speed $v$ for two angles of incidence, (a) $\theta =
90^\circ$ and (b) $\theta = 45^\circ$, and for several values of $r_{\rm min}$. For the smallest $r_{\rm min}$, the energy loss
presents a large maximum around the speed of $v=2.5$~a.u., and a smaller peak at a speed $v<1$~a.u. The physical reason for the
appearance of two peaks is likely due to the relatively broad gap between the low-energy and high-energy groups of peaks seen in
Fig.~\ref{fig:6}. As expected, the two peaks in Fig.~\ref{fig:7} decrease in magnitude as the minimum distance $r_{\rm min}$
increases, and they broaden and move towards higher speeds, so that the smaller peak turns into a shoulder around $v\sim
1-2$~a.u.. It is also interesting to notice that there is a threshold for energy loss on the order of $v\sim$ 0.3 a.u., which is
remarkably independent of the angle of incidence in Fig.~\ref{fig:7}. While such a threshold could not be anticipated from the
behavior of the average number of the lowest-energy $\pi$ plasmon mode $j=2$, $m=0$ in Fig.~\ref{fig:4}, it is perhaps
noteworthy that this threshold speed correlates with the slope of the dispersion curve for the mode $j=2$, $m=0$ when $k\gtrsim$
0.1 \AA$^{-1}$ in Fig.~\ref{fig:4}.

In the limit of grazing incidence, $\theta=0$, the particle travels parallel to the nanotube at a constant radial distance
$r_{\rm min}$. In that case, we can consider both internal ($r_{\rm min}<R$) and external ($r_{\rm min}>R$) particle
trajectories and define the stopping force acting on the particle, as described in the last paragraph of appendix~\ref{app:b}.
Figure \ref{fig:8} shows the results for stopping force at different positions (a) inside and (b) outside a SWCNT. We observe
that this force exhibits a peak structure which is quite similar to that observed in Fig.~\ref{fig:7} for the total energy loss,
with similar trends as the particle moves away from the nanotube wall. This can be rationalized by the fact that the shapes of
the curves for energy loss in Fig.~\ref{fig:7} are largely independent of the angle of incidence, and by considering the
stopping force as the energy loss of a particle per unit path length in the limiting case of oblique incidence when $\theta\to
0^\circ$.

\section{Conclusions}

The formulation developed in this work represents a direct but more realistic continuation of the works presented previously by
us~\cite{Mowbray_2006,Arista_2001} related to plasmon excitation by external charges moving paraxially in hollow cylindrical
nano-structures. We applied a quantization procedure~\cite{Arista_2001} to the two-fluid hydrodynamic model developed by Mowbray
et al.~\cite{Mowbray_2004}, in order to obtain the average number of plasmons excited, and the total energy loss suffered by a
fast charged particle passing near the surface of a single-walled carbon nanotube with an arbitrary angle of incidence.

One of the most important results of this version of the hydrodynamic model  is due to the inclusion of the restoring
interaction, which causes the $\sigma+\pi$ electron collective mode to oscillate at finite frequency in the limit of vanishing
wavenumber, and changes the quasi-acoustic dispersion of the $\pi$ plasmon so that it agrees well with the available EELS data.
On the other hand, quantization of these modes allowed us to obtain several quantities in terms of the average number of excited
plasmons, such as the stopping force, energy loss spectra and total energy loss. We studied these quantities as functions of
various parameters: the total speed of the incident particle, its minimum distance to the nanotube surface, and the inclination
angle of its trajectory with respect to the nanotube axis.

In addition, we have discussed (in appendix~\ref{app:b}) a relation between the quantized and semi-classical approaches. The
latter approach presents the possibility of including the effects of plasmon damping and hence enables a more direct comparison
with the available experimental energy loss data, which is left for future work.

\acknowledgments This work was supported by the Natural Sciences and Engineering Research Council of Canada. Financial support
from Agencia Nacional de Promoci\'on Cient\'\i fica y Tecnol\'ogica (PICT 32983) and Universidad Nacional de Cuyo is gratefully
acknowledged. \clearpage

\appendix

\section{Statistics of plasmon excitations on carbon nanotubes}
\label{app:a} In this appendix we use the methods of classical statistics to discuss energy losses of a fast charged particle
due to electron plasma excitations on a carbon nanotube. We have seen that the quantization approach in the main text represents
plasma excitations as a pool of independent quantum oscillators, or plasmons of mode $\qb=(j,m,k)$ with an eigenfrequency
$\omega_\qb=\omega_{jmk}$.

According to Carruthers and Nieto~\cite{Carruthers_1965}, the probability of exciting $N$ plasmons of any given mode $\qb$ is
given by a classical Poissonian distribution,
\begin{equation}
  \label{eq:b1}
  P_\qb(N)=\exponent^{-\mu_\qb}\,\frac{\mu_\qb^N}{N!},
\end{equation}
where $\mu_\qb = \mu_{jmk}$ is the mean number of plasmons excited in mode $\qb$. For further consideration, it is convenient to
define the characteristic (or moment-generating) function associated with the probability $P_{\qb}(N)$,
\begin{eqnarray}
  \label{eq:b3}
  G_{\qb}(\zeta)
  \equiv \langle \exponent^{i\zeta N}\rangle_{\qb}
  =\sum_{N=0}^{\infty}\, \exponent^{iN\zeta}\,P_{\qb}(N)
  =\exponent^{\mu_\qb(\exponent^{i\zeta}-1)}.
\end{eqnarray}

It is sometimes of practical or theoretical interest to discuss marginal probability distribution $P_{jm}(N)$ for having $N$
plasmons excited in modes with given $j$ and $m$ values, while the longitudinal wavenumber $k$ takes a full range of allowed
values. Because of the statistical independence of plasmons of mode $\qb$, the excitation of plasmons with fixed $(j,m)$ can be
imagined as a sub-ensemble consisting of independent oscillators with different values of $k$. Then, the characteristic function
associated with the marginal probability $P_{jm}(N)$ can be written as a product of the characteristic functions
$G_{\qb}(\zeta)$ for each member of this sub-ensemble,
\begin{equation}
  \label{eq:b4}
  G_{jm}(\zeta) = \prod_k G_{\qb}(\zeta)
  =\prod_k
      \exponent^{\mu_{\qb}(\exponent^{i\zeta}-1)}=\exponent^{\mu_{jm}(\exponent^{i\zeta}-1)},
\end{equation}
where $\mu_{jm}\equiv\sum_k\mu_{\qb}$. By expanding the final result in Eq.~(\ref{eq:b4}) in a series of powers of the factor
$\exponent^{i\zeta}$, we find that the probability of exciting $N$ plasmons in the mode $(j,m)$ is also a Poissonian
distribution,
\begin{equation}
  \label{eq:b7}
  P_{jm}(N)=\exponent^{-\mu_{jm}}\,\frac{\mu_{jm}^N}{N!},
\end{equation}
with the average number of plasmons, $\mu_{jm}$, given in Eq.~(\ref{eq:39}).

Further, assuming that $N_\qb$ is the number of plasmons excited in the mode $\qb$ with frequency $\omega_\qb$, one can express
the probability density $P(\vareps)$ for losing energy $\varepsilon$ as an ensemble average taking into account the above
mentioned statistical independence of plasmons of mode $\qb$,
\begin{eqnarray}
  \label{eq:b8}
  P(\vareps) &=& \left<\delta\left(\vareps-\hslash\sum_{\qb}N_\qb
    \omega_\qb\right)\right>
  \nonumber \\
  &=& \int_{-\infty}^\infty
  \frac{d\tau}{2\pi\hslash}\,\exponent^{-i\vareps\tau/\hslash}\,\prod_\qb
 \left<\exponent^{iN_\qb\omega_\qb
       \tau}\right>_\qb
         \nonumber \\
 &\equiv& \int_{-\infty}^\infty\frac{d\tau}{2\pi\hslash}\,
\exp\left[-i\frac{\vareps\tau}{\hslash}
   +\sum_\qb \mu_\qb (\exponent^{i\omega_\qb \tau}-1)\right],
\end{eqnarray}
where the last step was derived by the use of Eq.~(\ref{eq:b3}). By expanding the final result in Eq.~(\ref{eq:b8}) in a power
series of the factor $\exponent^{i\omega_\qb \tau}$, we obtain the probability density as
\begin{eqnarray}
  \label{eq:b16}
  P(\vareps) = \sum_{\qb}\mu_\qb\delta(\vareps-\hslash\omega_\qb).
\end{eqnarray}
This can be written in a more explicit form by invoking Eq.~(\ref{eq:35}) and using the delta function in Eq.~(\ref{eq:b16}) to
replace $\omega_\qb\equiv\omega_{jmk}\rightarrow\vareps/\hslash$, as
\begin{eqnarray}
  \label{eq:b17}
  P(\vareps)  & = &
   \sum_{jmk}\delta(\vareps-\hslash\omega_{jmk})\,C_{jmk}
  \left|\VV\left(R;m,k,\frac{\vareps}{\hslash}\right)\right|^2
   \nonumber \\
   & \rightarrow & \sum_{m=-\infty}^\infty\frac{L}{2\pi}\int_{-\infty}^\infty dk\,
   \left|\VV\left(R;m,k,\frac{\vareps}{\hslash}\right)\right|^2\sum_{j=1}^2 C_{jmk}
   \delta\left(\vareps-\hslash\omega_{jmk}\right),
 \end{eqnarray}
where $C_{jmk}$ is given in Eq.~(\ref{eq:36}), whereas $\VV(R;m,k,\omega)$ is given in Eqs.~(\ref{eq:37a}), (\ref{eq:37}), and
(\ref{eq:38}) for a straight line trajectory.

Finally, we note that the above result for probability density of energy loss can be used to evaluate the total energy loss as
the mean value, $\Eloss=\int\!d\vareps \, \vareps\, P(\vareps) $, so that
\begin{eqnarray}
  \label{eq:b10}
  \Eloss = \sum_\qb \hslash \omega_\qb \mu_\qb \rightarrow
  \sum_{j=1}^2\sum_{m=-\infty}^\infty
  \frac{\hslash L}{2\pi}\int_{-\infty}^\infty dk\,\omega_{jmk}\mu_{jmk}.
\end{eqnarray}
It will be shown in the appendix~\ref{app:b} that the above results for $P(\vareps)$  and $\Eloss$ have a close relation with a
semi-classical approach based on the hydrodynamic model.

\section{Semi-classical model of plasmon excitations on carbon nanotube}
\label{app:b} A semi-classical, two-fluid 2D hydrodynamic model of plasmon excitations on a carbon nanotube can be obtained by
using the equation of motion Eq.~(\ref{eq:2}) for the function $\xi_\nu(\rbs,t)\equiv\xi_\nu(\varphi,z,t)$ augmented by the
friction,
\begin{eqnarray}
 \label{eq:c1}
m_\nu^*\left[\ddotxi_\nu+\eta_\nu\dotxi_\nu+\omega_{\nu
  r}^2\xi_\nu-s_\nu^2\grad^2\xi_\nu\right]=\Vext(\rbs,t)-\charge\Phi_\text{ind}(\rbs,t),
\end{eqnarray}
where $\eta_\nu>0$ are friction coefficients for the $\nu$th fluid, which can be used to describe plasmon damping in a
phenomenological manner.

The system of equations Eq.~(\ref{eq:c1}), coupled with the Poisson equation Eq.~(\ref{eq:5}) for the induced potential on the
nanotube surface, can be solved by using the Fourier transform with respect to coordinates and time, $\{\varphi,z,t\}\rightarrow
\{m,k,\omega\}$, so that
\begin{eqnarray}
\label{eq:c2} \breve{\xi}_\nu(m,k,\omega)= \int_{-\infty}^\infty dt\, \exponent^{i\omega t} \xfb_\nu(m,k,t),
\end{eqnarray}
with $\xfb_\nu(m,k,t)$ defined in Eq.~(\ref{eq:6}). Using the fact that the total induced number density of electrons per unit
area is $\deltan=n^0_\sigma\grad^2\xi_\sigma+n^0_\pi\grad^2\xi_\pi$, we can write its Fourier transform as
\begin{eqnarray}
 \label{eq:c4}
 \deltaFn(m,k,\omega)=-\chi(m,k,\omega)\VV(R;m,k,\omega),
\end{eqnarray}
where the density response function of a carbon nanotube is given by
\begin{eqnarray}
\label{eq:c5} \chi(m,k,\omega)=\frac{\chi_0(m,k,\omega)}{1+4\pi \charge^2R I_m(|k|R)K_m(|k|R)\chi_0(m,k,\omega)},
\end{eqnarray}
with $\chi_0=\chi_\sigma^{(0)}+\chi_\pi^{(0)}$, where the non-interacting response function of the $\nu$th fluid is given by
\begin{eqnarray}
\label{eq:c6}
\chi_\nu^{(0)}(m,k,\omega)=\frac{\frac{n_\nu^0}{m_\nu^*}\left(k^2+\frac{m^2}{R^2}\right)}{s_\nu^2\left(k^2+\frac{m^2}{R^2}\right)+\omega_{\nu
r}^2-\omega\left(\omega+i\eta_\nu\right)}.
\end{eqnarray}
It may be worthwhile quoting the final expression for $\chi(m,k,\omega)$ in terms of the quantities defined in
Eqs.~(\ref{eq:13}), (\ref{eq:14}), and (\ref{eq:20}) and finite friction coefficients,
\begin{eqnarray}
\label{eq:c14a} \chi(m,k,\omega)=\frac{ \left(\omega_\pi^2-i\omega\eta_\pi\right)\frac{n_\sigma^0}{m_\sigma^*} +
\left(\omega_\sigma^2-i\omega\eta_\sigma\right)\frac{n_\pi^0}{m_\pi^*} -\omega^2\frac{n_0}{m_*}
-\Delta^2\sqrt{\frac{n_\sigma^0}{m_\sigma^*}\frac{n_\pi^0}{m_\pi^*}} }{
 \left[\omega_\pi^2-\omega\left(\omega+i\eta_\pi\right)\right]
\left[\omega_\sigma^2-\omega\left(\omega+i\eta_\sigma\right)\right] -\Delta^4 } \left( k^2+\frac{m^2}{R^2}\right).
 \end{eqnarray}

We define the total energy loss as the work done by the induced force on the external charge as it moves along its entire
trajectory $\rb_0(t)$ with  velocity ${\bf v}_0(t)={\bf \dot{r}}_0(t)$,
\begin{eqnarray}
\label{eq:c8}
 \Eloss =-\int_{-\infty}^\infty dt \,{\bf v}_0(t)\cdot\vFind({\bf r}_0(t),t),
\end{eqnarray}
where
\begin{eqnarray}
\label{eq:c9} \vFind(\rb_0(t),t)=-\charge Z\left.\nabla\Phiind(\rb,t)\right|_{\rb = \rb_0(t)}.
\end{eqnarray}
On substituting Eq.~(\ref{eq:c9}) in Eq.~(\ref{eq:c8}) and using the chain rule, $\frac{d}{dt} = {\bf v}_0(t)\cdot\nabla +
\frac{\partial}{\partial t}$, we can eliminate the conservative part of the time integral in Eq.~(\ref{eq:c8}), thus arriving at
\begin{eqnarray}
\label{eq:c10} \Eloss&=&-\charge Z\int_{-\infty}^\infty dt\, \left[\frac{\partial}{\partial t} \Phiind(\rb,t) \right]_{\rb =
\rb_0(t)}
\nonumber\\
&=& \int_{-\infty}^\infty \frac{d\omega}{2\pi i}\,\omega \int_{-\infty}^\infty dt\, \exponent^{-i\omega t}
\FTPhiind(\rb_0(t),\omega),
\end{eqnarray}
where $\FTPhiind(\rb,\omega)$ is the Fourier transform of the induced potential with respect to time only,
\begin{eqnarray}
\label{eq:c11} \FTPhiind(\rb,\omega)= \int_{-\infty}^\infty d\omega \,\exponent^{i\omega t} \Phiind(\rb, t),
\end{eqnarray}
which can be obtained at arbitrary points outside the nanotube from
\begin{eqnarray}
\label{eq:c7} \Phiind(\rb,t)=\frac{-\charge}{2\pi
}\sum_{m=-\infty}^{\infty}\int_{-\infty}^{\infty}\frac{dk}{2\pi}\exponent^{im\varphi+ikz} g_{mk}(r,R)\,\deltaFn(m,k,t),
\end{eqnarray}
with $ g_{mk}(r,R)=4\pi I_m(|k|R)K_m(|k|r)$.

Finally, by combining Eqs.~(\ref{eq:c4}), (\ref{eq:c10}), (\ref{eq:c11}), and (\ref{eq:c7}), and referring to the definitions in
Eqs.~(\ref{eq:33}) and (\ref{eq:37a}), we can write the total energy loss as
\begin{eqnarray}
\label{eq:c12} \Eloss=\frac{1}{\pi R}\sum_{m=-\infty}^\infty\int_{-\infty}^\infty \frac{dk}{(2\pi)^2}\int_{0}^\infty
d\omega\,\omega
   \left|\VV\left(R;m,k,\omega\right)\right|^2\Im\left[\chi(m,k,\omega)\right],
 \end{eqnarray}
where we have used the property that the real and imaginary parts of the density response, $\chi(m,k,\omega)$, are an even and
an odd function of frequency $\omega$, respectively. Since the total energy loss can be written as a first moment of a
semi-classical version of the probability density for energy loss $\vareps=\hslash\omega$,
\begin{eqnarray}
\label{eq:c13} \Eloss=\int_{0}^\infty d\vareps \, \vareps P_\text{sc}(\vareps),
 \end{eqnarray}
by comparison with Eq.~(\ref{eq:c12}), one can deduce
\begin{eqnarray}
\label{eq:c14} P_\text{sc}(\vareps)=\frac{1}{\pi R}\sum_{m=-\infty}^\infty\int_{-\infty}^\infty \frac{dk}{(2\pi\hslash)^2}
   \left|\VV\left(R;m,k,\frac{\vareps}{\hslash}\right)\right|^2\Im\left[\chi\left(m,k,\frac{\vareps}{\hslash}\right)\right].
 \end{eqnarray}

It can be shown that the expression in Eq.~(\ref{eq:c14}) reduces to the limiting expression in Eq.~(\ref{eq:b17}), which is
obtained using the quantization of plasma excitations, i.e., when the plasmon damping is neglected. Namely, when both friction
coefficients in the semi-classical hydrodynamic model are vanishingly small, $\eta_\nu\rightarrow 0^+$, we obtain for the
imaginary part of the nanotube's response function at positive frequencies, $\omega>0$,
\begin{eqnarray}
  \label{eq:c15}
 \Im\left[\chi(m,k,\omega)\right]=\frac{\pi}{2}\frac{n_0}{m_*}\frac{\frac{m^2}{R^2}+k^2}{\omega}
 \sum_{j=1}^2D^2_{jmk}\,\delta\left(\omega-\omega_{jmk}\right),
\end{eqnarray}
with the coefficient $D_{jmk}$ defined in Eqs.~(\ref{eq:22}), (\ref{eq:24}) and (\ref{eq:23}). Using Eq.~(\ref{eq:36}), this can
be written as $\Im\left[\chi(m,k,\omega)\right]=\pi\hslash\mathcal{A}\sum_jC_{jmk}\,\delta\left(\omega-\omega_{jmk}\right)$,
which, when substituted in Eq.~(\ref{eq:c14}), gives the probability density of energy loss Eq.~(\ref{eq:b17}) of the
quantization approach. Of course, unlike the quantization approach, it is possible, and even desirable to study the effects of
plasmon damping on energy-loss distributions by using finite values for $\eta_\nu$ in the present semi-classical approach.

Moreover, the result in Eq.~(\ref{eq:c15}) can also be used to show that the semi-classical result for the total energy loss,
Eq.~(\ref{eq:c12}), is equivalent to the result Eq.~(\ref{eq:b10}) of the plasma quantization approach, in the limit of
vanishing plasmon damping. Of particular interest here is the case when the external perturbing charge moves parallel to a
nanotube with speed $v_\parallel$ (while $v_\perp=0$) at fixed distance $r_0>R$, so that $\varphi_0=0$ and $z_0(t)=v_\parallel
t$. We initially assume that the nanotube has a finite length $L$, and hence the traversal time of the external particle is
$T=L/v_\parallel$. Using the expressions Eqs.~(\ref{eq:33}) and (\ref{eq:37a}), we obtain in the limit $L\rightarrow\infty$,
\begin{eqnarray}
  \label{eq:c16}
\left|\VV\left(R;m,k,\omega\right)\right|^2 &=&
 \left[\charge^2ZRg_{mk}(R,r_0)\right]^2\left|\int_{-T/2}^{T/2}dt\,\exponent^{i\left(\omega-kv_\parallel\right)t}\right|^2
\nonumber\\
&\approx& 2\pi\frac{L}{v_\parallel}\left[\charge^2ZRg_{mk}(R,r_0)\right]^2
 \,\delta\left(\omega-kv_\parallel\right),
\end{eqnarray}
allowing us to derive from Eqs.~(\ref{eq:c12}) the average stopping force for the external particle, $S=\Eloss/L$, as
\begin{eqnarray}
  \label{eq:c17}
S=8R(\charge^2Z)^2 \sum_{m}\int_{0}^\infty dk\,k\,I_m^2(|k|R)K_m^2(|k|r_0)\Im\left[\chi(m,k,kv_\parallel)\right],
\end{eqnarray}
when $r_0 > R$, as shown in Fig.~\ref{fig:8}(b).  On the other hand, when the perturbing charge is inside the SWNT, $r_0 < R$,
we obtain an analogous expression for the stopping force of
\begin{eqnarray}
\label{eq:c18}
S=8R(\charge^2Z)^2 \sum_{m}\int_{0}^\infty dk\,k\,I_m^2(|k|r_0)K_m^2(|k|R)\Im\left[\chi(m,k,kv_\parallel)\right],
\end{eqnarray}
as shown in Fig.~\ref{fig:8}(a). It should be noted that these expressions for the stopping force are identical to those
previously derived from semi-classical models of carbon nanotubes \cite{Mowbray_2004,Mowbray_thesis}.

\newpage
\section{Planar case}
\label{app:c} We outline here how the formalism of the main text can be adapted to describe plasmon excitation in a planar 2D
electron gas by an external charge moving on a specularly reflected trajectory based on the two-fluid model. There may be some
interest for this outline in view of possible applications to EELS experiments on free-standing graphene
\cite{Eberlein_2008,Liu_2008}.

The planar case is simply retrieved by assuming that the nanotube radius grows indefinitely, $R\rightarrow\infty$, in such a
manner that the position along the nanotube circumference can be defined by a Cartesian coordinate $y=R\varphi$, whereas the
radial distance from the nanotube wall can be defined by the Cartesian coordinate $x=r-R$. Similarly, with the longitudinal
wavenumber $k$ renamed $k_z$, the ratio $m/R$ becomes a quasi-continuous variable, $k_y$, corresponding to a wavenumber for
collective modes propagating around the nanotube's circumference. Consequently, the $d\varphi$ integral and the $m$ summation in
Eqs.\ (\ref{eq:6}) and (\ref{eq:7}) become, respectively,
\begin{eqnarray}
  \label{eq:d1}
R\int\limits_{-\pi}^\pi d\varphi\cdots &\rightarrow &\int\limits_{-\infty}^\infty dy\cdots,
\\
  \label{eq:d2}
 \frac{1}{2\pi R}\sum_{m=-\infty}^\infty\cdots & \rightarrow & \int\limits_{-\infty}^\infty \frac{dk_y}{2\pi}\cdots,
\end{eqnarray}
indicating that we are using a 2D Fourier transform in the plane of the electron gas that maps
$\{y,z\}\rightarrow\{k_y,k_z\}\equiv\Kb$. Accordingly, the generic expression $m^2/R^2+k^2$ appearing in the equations of the
main text is mapped to $K^2=k_y^2+k_z^2$, and the Green function becomes
\begin{eqnarray}
  \label{eq:d3}
  R\,g_{mk}(R,r)
  \rightarrow  \frac{2\pi}{K}\mathrm{e}^{-K\vert x\vert},
\end{eqnarray}
to the leading order in $1/R$.

Next, we define a specularly reflected trajectory for the incident particle in Cartesian coordinates by
$\rb_0(t)=\{x_0(t),y_0(t),z_0(t)\}=\{r_0(t)-R,R\varphi_0(t),z_0(t)\}\rightarrow\{x_\mathrm{min}+V_\perp\vert t\vert,V_{\parallel
y}t,V_{\parallel z}t\}$, where $x_\mathrm{min}>0$ is the minimum distance from the planar electron gas and $V_\perp$ is the
normal component of the projectile velocity. Then, Eq.\ (\ref{eq:33}) gives in the limit $R\to\infty$
\begin{equation}
  \label{eq:d4}
  \widetilde{V}_{\rm ext} (R;m,k,t) \rightarrow  -e^2Z\,
\frac{2\pi}{K}\mathrm{e}^{-K\left(x_\mathrm{min}+V_\perp\vert t\vert\right)-i\Kb\cdot\Vpb t},
\end{equation}
where $\Vpb=\{V_{\parallel y},V_{\parallel z}\}$ is the parallel component of the projectile velocity. As a consequence, Eq.\
(\ref{eq:37a}) gives
\begin{equation}
  \label{eq:d5}
  \VV(R;m,k,\omega)\rightarrow
  -e^2Z\,
\frac{4\pi}{K}\mathrm{e}^{-K x_\mathrm{min}}\frac{KV_\perp}{\left(\omega-\Kb\cdot\Vpb\right)^2+\left(KV_\perp\right)^2},
\end{equation}
so that the average number of plasmons in the mode $(j,m,k)\rightarrow (j,\Kb)$ finally follows from Eq.\ (\ref{eq:35}) as
\begin{equation}
  \label{eq:d6}
\mu_{jmk}\rightarrow \frac{2n_0}{\hbar m_*}\,\frac{D_{j\Kb}^2}{\omega_{j\Kb}}\left(e^2Z\right)^2\,\mathrm{e}^{-2K
x_\mathrm{min}}\,\frac{\left(KV_\perp\right)^2}{\left[\left(\omega_{j\Kb}-\Kb\cdot\Vpb\right)^2+\left(KV_\perp\right)^2\right]^2},
\end{equation}
where $D_{j\Kb}$ is given by Eqs.\ (\ref{eq:22}), (\ref{eq:24}) and (\ref{eq:23}), in which plasmon frequencies are to be used
from Eqs.\ (\ref{eq:13}), (\ref{eq:14}) and (\ref{eq:18}) upon the replacements
\begin{equation}
  \label{eq:d7}
\omega_\nu^2\rightarrow\omega_{\nu r}^2+2\pi e^2\frac{n_\nu^0}{m_\nu^*}K+s_\nu^2K^2,
\end{equation}
and
\begin{equation}
 \label{eq:d8}
\Delta^2\rightarrow 2\pi e^2\sqrt{\frac{n_\sigma^0}{m_\sigma^*}\frac{n_\pi^0}{m_\pi^*}}\,K.
\end{equation}

It is worth mentioning that, if one lets both $\omega_{\nu r}\to 0$ and $s_\nu\to 0$ in Eq.\ (\ref{eq:d7}), the planar version
of Eq.\ (\ref{eq:18}) no longer gives rise to plasmon splitting, but rather one recovers $\omega_{2\Kb}\to 0$ and
$\omega_{1\Kb}\to\sqrt{2\pi e^2\frac{n_0}{m_*}K}$ describing the familiar plasmon dispersion of a single-fluid model for planar
2D electron gas with surface density $n_0=n_\sigma^0+n_\pi^0$ and an effective mass $m_*$ \cite{Fetter_1973}.

\end{document}